\documentclass[apj]{emulateapj}

\usepackage{graphicx} 
\usepackage{amsmath}
\usepackage{appendix}
\usepackage{amssymb}
\usepackage{latexsym}
\usepackage{longtable}
\usepackage{threeparttable} 
\usepackage{threeparttablex} 
\usepackage{subfigure} 
\usepackage[figoff]{figcaps}
\usepackage{ccaption}
\captiondelim{ } 

\begin{document}

\pagestyle{plain}
\title{Orbits of Near-Earth Asteroid Triples 2001 SN263 and 1994 CC: \\ Properties, Origin, and Evolution}
\slugcomment{Accepted to Astronomical Journal}
\author{Julia Fang\altaffilmark{1}, Jean-Luc Margot\altaffilmark{1,2}, Marina Brozovic\altaffilmark{3}, Michael C. Nolan\altaffilmark{4},\\ Lance A. M. Benner\altaffilmark{3}, Patrick A. Taylor\altaffilmark{4}}

\altaffiltext{1}{Department of Physics \& Astronomy, University of California Los Angeles, Los Angeles, CA 90095, USA}
\altaffiltext{2}{Department of Earth and Space Sciences, University of California Los Angeles, Los Angeles, CA 90095, USA}
\altaffiltext{3}{Jet Propulsion Laboratory, California Institute of Technology, Pasadena, CA 91109, USA}
\altaffiltext{4}{Arecibo Observatory, National Astronomy and Ionosphere Center, Arecibo, PR 00612, USA}

\begin{abstract}
Three-body model fits to Arecibo and Goldstone radar data reveal the nature of two near-Earth asteroid triples. Triple-asteroid system 2001 SN263 is characterized by a primary of $\sim$10$^{13}$ kg, an inner satellite $\sim$1$\%$ as massive orbiting at $\sim$3 primary radii in $\sim$0.7 days, and an outer satellite $\sim$2.5$\%$ as massive orbiting at $\sim$13 primary radii in $\sim$6.2 days. 1994 CC is a smaller system with a primary of mass $\sim$2.6 $\times 10^{11}$ kg and two satellites $\sim$2$\%$ and $\lesssim$$1\%$ as massive orbiting at distances of $\sim$5.5 and $\sim$19.5 primary radii. Their orbital periods are $\sim$1.2 and $\sim$8.4 days. Examination of resonant arguments shows that the satellites are not currently in a mean-motion resonance. Precession of the apses and nodes are detected in both systems (2001 SN263 inner body: $d\varpi/dt \sim$1.1 deg/day, 1994 CC inner body: $d\varpi/dt \sim$ -0.2 deg/day), which is in agreement with analytical predictions of the secular evolution due to mutually interacting orbits and primary oblateness. Nonzero mutual inclinations between the orbital planes of the satellites provide the best fits to the data in both systems (2001 SN263: $\sim$14 degrees, 1994 CC: $\sim$16 degrees). Our best-fit orbits are consistent with nearly circular motion, except for 1994 CC's outer satellite which has an eccentric orbit of $e \sim$ 0.19. We examine several processes that can generate the observed eccentricity and inclinations, including the Kozai and evection resonances, past mean-motion resonance crossings, and close encounters with terrestrial planets. In particular, we find that close planetary encounters can easily excite the eccentricities and mutual inclinations of the satellites' orbits to the currently observed values.
\end{abstract}
\maketitle

\section{Introduction}

The existence and prevalence ($\sim$16\%) of binary asteroids in the near-Earth population (Margot et al. 2002; Pravec et al. 2006) naturally lead to the search and study of multiple-asteroid systems (Merline et al. 2002; Noll et al. 2008). Triple systems are known to exist in the outer Solar System, the main belt, and the near-Earth population. Among the trans-neptunian objects (TNOs), there are currently two well-established triples, 1999 TC36 (Margot et al. 2005; Benecchi et al. 2010) and Haumea (Brown et al. 2006), and one known quadruple, Pluto/Charon (Weaver et al. 2006). In the main belt population, four triples are known to exist: 87 Sylvia, 45 Eugenia, 216 Kleopatra, and 3749 Balam (Marchis et al. 2005; Marchis et al. 2007; Marchis et al. 2008a; Marchis et al. 2008b). There are currently only two well-established asteroid triples in the near-Earth population, 2001 SN263 (Nolan et al. 2008a) and 1994 CC (Brozovic et al. 2009), both of which are the focus of this study. There is also another possible triple, near-Earth asteroid 2002 CE26, that may have a tertiary component but the limited observational span of this object prevented an undisputable detection (Shepard et al. 2006).

\begin{figure}[htb]
	\centering
	\includegraphics[scale=0.45]{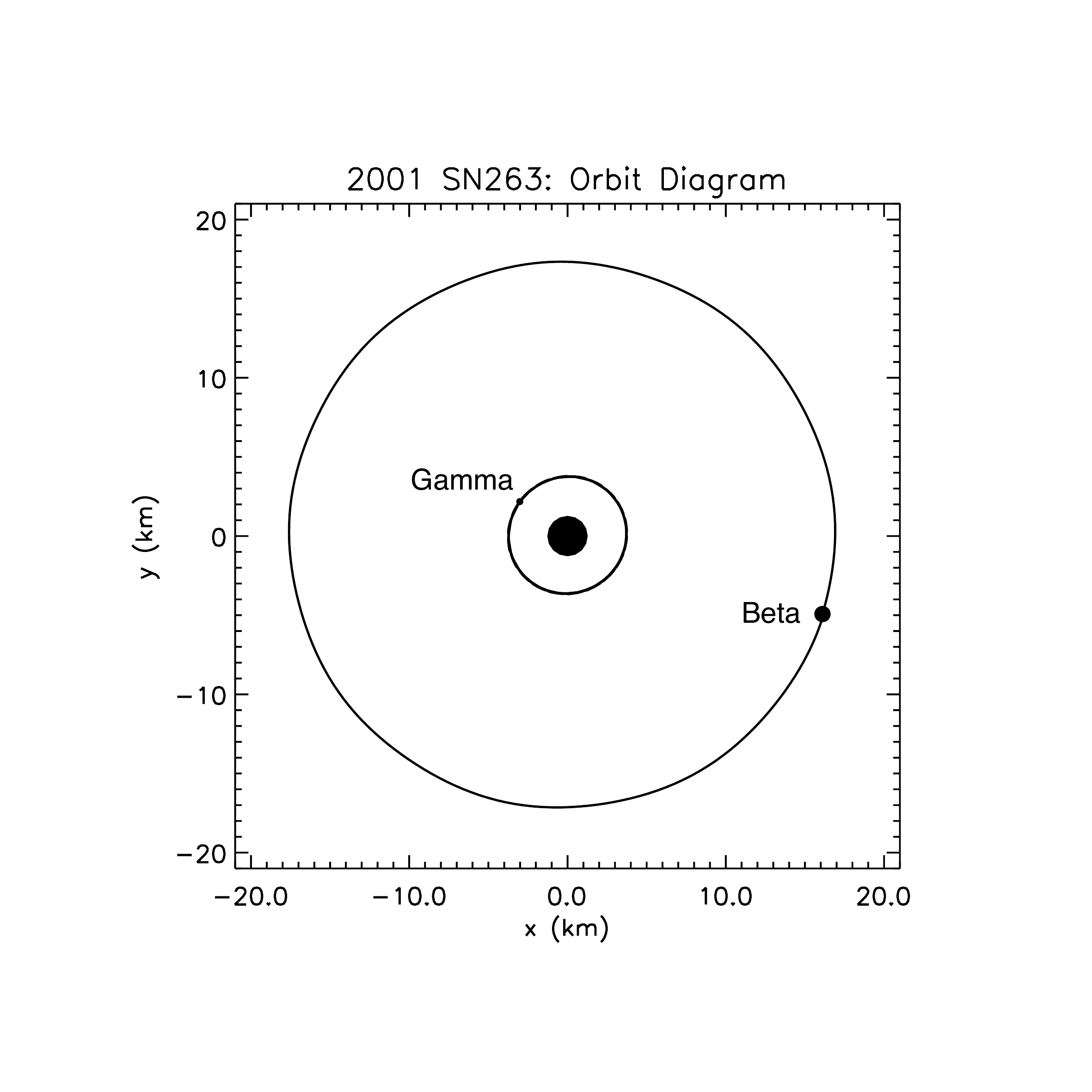}
	\includegraphics[scale=0.45]{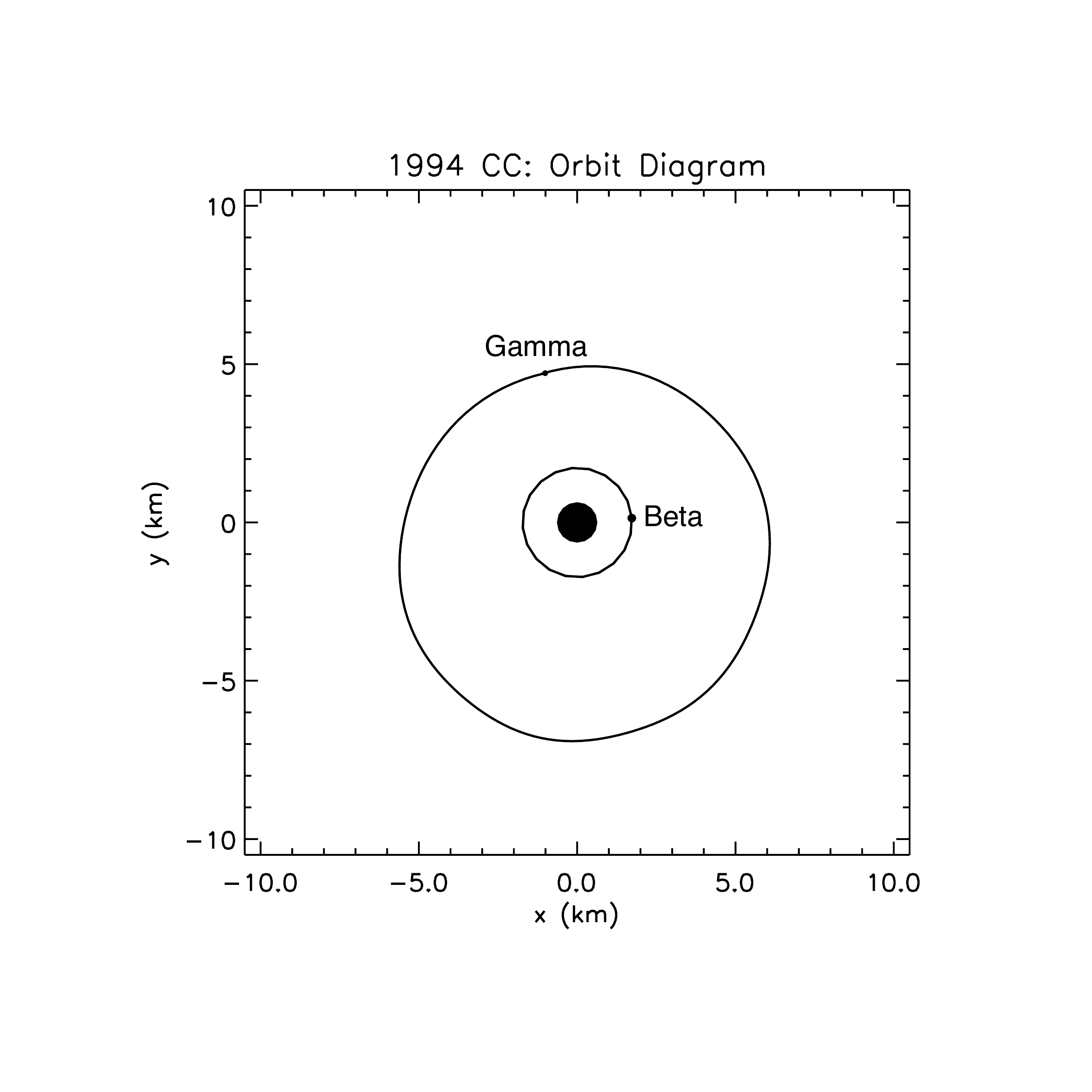}
	\caption{2001 SN263 and 1994 CC: Best-fit orbit diagrams of the inner and outer satellites, projected onto Alpha's equatorial plane. In both diagrams, we show the actual trajectories from numerical integrations, and the relative sizes of the bodies (estimated from radar images) are shown to scale. All bodies are located at their positions at MJD 54509 (for 2001 SN263) and MJD 54994 (for 1994 CC) with Alpha centered on the origin. The slightly irregular shape of the orbit of 1994 CC's outer body is real and due to mutual perturbations in the system. \label{diagrams}} 
\end{figure}

(153591) 2001 SN263 has been unambiguously identified as a triple-asteroid system, and its orbit around the Sun is eccentric at 0.48 with a semi-major axis of 1.99 AU and inclined 6.7 degrees with respect to the ecliptic. It is an Amor asteroid with a pericenter distance of 1.04 AU. The system is composed of three components: a central body (equivalent radius $\sim$ 1.3 km) and two orbiting satellites. In this paper, we use the following terminology for triple systems: the central body (most massive component) is called Alpha, the second most massive body is termed Beta, and the least massive body is named Gamma (Figure \ref{diagrams}). In the case of 2001 SN263, Beta is the outer satellite and Gamma is the inner satellite. The other near-Earth triple is (136617) 1994 CC with a primary of equivalent radius R $\sim$ 315 m, where the opposite is true; Beta is the inner body and Gamma is the outer body. Its heliocentric orbit has a semi-major axis of 1.64 AU and is also eccentric (0.42) and inclined (4.7 degrees) with respect to the ecliptic. 1994 CC is an Apollo asteroid with a pericenter distance of 0.95 AU.

In this work, we present dynamical solutions for both triple systems, 2001 SN263 and 1994 CC, where we derived the orbits, masses, and Alpha's $J_2$ gravitational harmonic using N-body integrations. We utilize range and Doppler data from Arecibo and Goldstone, and these observations as well as our methods are described in Section 2. In Section 3, we present our best orbital solutions and their uncertainties. We also include discussion regarding the satellite masses and the primary's oblateness (described by $J_2$), and the observed precession of the apses and nodes are compared to analytical predictions. Section 4 describes the origin and evolution of the orbital configurations, including Kozai and evection resonant interactions, as well as the effects of planetary encounters.

Previous studies of multiple TNO systems include the analytic theory of Lee \& Peale (2006) for Pluto, where they treated Nix and Hydra as test particles. Tholen et al. (2008) used four-body orbit solutions to constrain the masses of Nix and Hydra; they did not find evidence of mean-motion resonances in the system. Ragozzine \& Brown (2009) determined the orbits and masses of Haumea's satellites using astrometry from Hubble Space Telescope and the W. M. Keck Telescope. They used a three-body model and found that their data was not sufficient to constrain the oblateness, described by $J_2$, of the non-spherical central body. Their orbital solutions yielded a large eccentricity ($\sim$0.249) of the inner, fainter satellite, Namaka, and a mutual inclination with the outer satellite, Hi'iaka, of $\sim$13.41 degrees. They postulated that the excited state of the system could be conceptually explained by the satellites' tidal evolution through mean-motion resonances. In the main belt, Winter et al. (2009) studied the orbital stability of the satellites in the Sylvia triple system and Marchis et al. (2010) presented a dynamical solution of Eugenia and its two satellites.

\section{Observations and Methods}

	To arrive at three-body orbit solutions for 2001 SN263 and 1994 CC, we used radar observations from Arecibo and Goldstone. Specifically, these observables include the range and Doppler separations between Alpha and Beta (or Gamma) at multiple epochs. Range-Doppler separations were measured as the center of mass (COM) differences between Beta (or Gamma) and Alpha. For 2001 SN263, we modeled the shapes of the components (Hudson 1993; Magri et al. 2007) and differenced the COM estimates. For 1994 CC, the COMs were estimated visually, using the center of the trailing edge of the echo as an estimate of the COM location, and taking range smear into account when necessary. Measurement uncertainties were based on the image resolution.  Our fits with reduced chi-square less than unity (see Section 3) indicate that uncertainties were assigned conservatively. For 2001 SN263, we used Arecibo observations taken over a span of approximately 14 days from the Modified Julian Date (MJD = JD - 2400000.5) 54508.06138 to MJD 54522.12406 when the triple made a close approach with Earth at 0.066 AU ($\sim$1550 R$_{\Earth}$). Range uncertainties are 75-150 meters and Doppler uncertainties are about 0.6 Hz at the nominal 2380 MHz frequency of the Arecibo radar (in this work, all Doppler observations and uncertainties are reported relative to this reference frequency). For 1994 CC, our observations were taken from both Arecibo and Goldstone planetary radars and spanned a total interval of almost 7 days from MJD 54994.69293 to MJD 55001.53906 during a close approach with Earth at 0.017 AU ($\sim$400 R$_{\Earth}$). The range uncertainties were 30-40 meters for Goldstone data and 25-75 meters for Arecibo data, and the Doppler uncertainties were 0.15-0.20 Hz for Goldstone data and 0.20-0.40 Hz for Arecibo data. In total, for 2001 SN263 we obtained 128 ranges and 128 Doppler measurements for each satellite, and for 1994 CC we had 112 ranges and 112 Doppler measurements for Beta and 78 ranges and 78 Doppler measurements for Gamma. Observations are available upon request.

Using these observations, the orbits, masses, and $J_2$ values of 2001 SN263 and 1994 CC were obtained using a model that includes the mutual gravitational interactions of all three bodies. To calculate the orbital evolution of the triples' components to fit to observations, we used a general Bulirsch-Stoer algorithm from an N-body integrator package called \verb MERCURY , version 6.2 (Chambers 1999). The Bulirsch-Stoer algorithm, although slow, was chosen for its computational accuracy. For our long-term integrations to test for stability, we used a fast hybrid symplectic/Bulirsch-Stoer algorithm, also part of the \verb MERCURY  package. Since the satellites for both systems are deep in the potential wells of their respective central bodies, we ignored perturbations from the Sun and planets during the orbit-fitting process. The Hill radius $r_{\rm Hill}$ of the central body (2001 SN263: $\sim$345 km, 1994 CC: $\sim$86 km), defined as the spherical region where the central body's gravity dominates the attraction of its satellites compared to the Sun, is much larger than the outer satellite's semi-major axis $a$ (2001 SN263: $\sim$16.6 km, 1994 CC: $\sim$6.1 km) in both systems. For 2001 SN263, $a_{\rm inner}/r_{\rm Hill}$ is 0.011 and $a_{\rm outer}/r_{\rm Hill}$ is 0.048. In the case of 1994 CC, $a_{\rm inner}/r_{\rm Hill}$ is 0.020 and $a_{\rm outer}/r_{\rm Hill}$ is 0.071. The assumption that we can ignore effects from external bodies breaks down during the long-term evolution of these triples and during close planetary encounters, which we address in Section 4.

\subsection{Short-Term Integrations: Orbit Fitting}

	Solving for the orbits, masses, and $J_2$ of the triple systems is a highly non-linear least-squares problem that requires 16 parameters, which we fitted simultaneously. This included two sets of six osculating orbital elements (semi-major axis, eccentricity, inclination, argument of pericenter, longitude of the ascending node, and mean anomaly at epoch), three masses, and a $J_2$ for the central body. To solve this minimization problem, we attempted to search for a global minimum using the Levenberg-Marquardt algorithm \verb mpfit  (Markwardt 2008), written in IDL. We ran \verb mpfit  with thousands of initial conditions to search for the best-fitting parameters. 
	
\def\arraystretch{1.4}
\begin{deluxetable}{l l l}
\tablecolumns{3}
\tablecaption{Orbit Model Constraints \label{constraints}}
\startdata
\hline \hline
& 2001 SN263 & 1994 CC \\
\hline
Central Mass (kg) & $10^{10}$ - none & $10^{10}$ - none \\
Beta (kg) & $10^9$ - none & $10^7$ - none \\
Gamma (kg) & $10^7$ - none & $10^6$ - none \\
$a$ (km) & 0 - none & 0 - none \\
$e$ & 0 - 0.9 & 0 - 0.9 \\
$J_2$ & 0 - 0.25 & 0 - 0.25
\enddata
\tablenotetext{}{These are a priori constraints that put limits on fitted parameters, so that a parameter can be bounded on the lower and/or upper side during the fitting process. The lower bounds on the masses were obtained by using size estimates from radar images and unit density, then dividing by a conservative factor of $\sim$100. In this table, `none' means that no limit was set and the parameter was allowed to float without limits in that direction.}
\end{deluxetable}

	For both systems, we constrained some parameters with upper and lower bounds (Table \ref{constraints}), and all others were allowed to float without restraint. The initial guesses for orbital parameters and masses were guided by hundreds of two-body Keplerian solutions, but substantially augmented from these to explore wide regions of parameter space including ``outlier" cases to be sure we covered all of the possibilities. Tables \ref{gridsn} and \ref{gridcc} show the ranges of starting guesses for our minimization routine as well as the maximum stepsize intervals between trial values. Typical stepsizes were significantly smaller than the maximum values listed. For the starting $J_2$ value of the central body, we adopted current estimates of axial ratios from shape modeling efforts (Nolan et al., in prep., Brozovic et al., in prep.) and a uniform density assumption. This resulted in initial $J_2$ values of 0.016 for 2001 SN263 and 0.013 for 1994 CC.
	
	The non-spherical nature of the central body (Alpha) introduces additional non-Keplerian effects. The distribution of mass within Alpha can be represented by the $J_n$ terms in its gravitational potential (Murray \& Dermott 1999), where the largest contribution is due to the lowest-order gravitational moment, the quadrupole term $J_2$ (there is no n=1 term since the origin of coordinates is Alpha's center of mass). $J_2$ is a good approximation for describing the oblateness of  primary bodies with substantial axial symmetry and as a result, we ignored higher-order terms. As we will see, even $J_2$ is not well-constrained, so there was no need to go to higher-order terms. It is related to three moments of inertia, A, B, and C, of the central body (Murray \& Dermott 1999):
\begin{equation}
		J_2 = \dfrac{C - \dfrac{1}{2}(A+B)}{MR^2} \approx \dfrac{C-A}{MR^2}
\end{equation}
	where $M$ is the total mass and $R$ is Alpha's equatorial radius. $J_2$ is an observable quantity because the oblateness of a body modifies the gravitational field experienced by orbiting satellites; in our case, the orbits of Beta and Gamma precessed through space as a consequence of Alpha's $J_2$. 

\def\arraystretch{1.4}
\begin{deluxetable}{l r r}
	\centering
	\tablecolumns{3}
	\tablecaption{2001 SN263: Range of Initial Conditions \label{gridsn}}
	\startdata
		\hline \hline
		 & Alpha & Maximum Stepsize \\
		\hline
		Mass ($10^{10}$ kg) & 850 - 980 & $\lesssim$20 \\
		$J_2$ & 0.0001 - 0.09 & $\lesssim$0.03 \\
		\hline
		 & Gamma (Inner) & Maximum Stepsize \\
		\hline
		Mass ($10^{10}$ kg) & 1 - 20 & $\lesssim$6\\
		$a$ (km) & 3.5 - 6.0 & $\lesssim$2.2 \\
		$e$ & 0 - 0.4 & $\lesssim$0.3 \\
		\hline
		& Beta (Outer) & Maximum Stepsize \\
		\hline
		Mass ($10^{10}$ kg) & 7 - 100 & $\lesssim$38 \\
		$a$ (km) & 16 - 24 & $\lesssim$6.8 \\
		$e$ & 0 - 0.4 & $\lesssim$0.3
		\enddata
	\tablenotetext{}{Ranges of initial conditions with the maximum stepsize, defined as the largest interval between two consecutive initial values. Most stepsizes were significantly smaller than those listed. All angles were examined over their full range with typical stepsizes of 5-10 degrees.}
\end{deluxetable}	

In most cases, we aligned Alpha's pole direction with the normal to Beta's orbital plane, consistent with the generally accepted formation process by spin-up and mass shedding (Margot et al. 2002; Pravec et al. 2006; Walsh et al. 2008). To be thorough, we also searched for situations where Alpha's pole might not be aligned with Beta's orbit normal by surveying a comprehensive range of right ascension (RA) and declination (DEC) values for Alpha's pole orientation. We also performed numerical integrations where we set the value of $J_2$ to fixed values (i.e. 0.005, 0.010, 0.015) such that it was no longer a floating parameter, which reduced the number of parameters to 15. In the particular case of 1994 CC, we also iterated through a list of pole orientations for Alpha suggested by the shape modeling process (Brozovic et al., in prep.). With these poles, we explored another vast grid of starting parameters. This resulted in thousands of additional fits; however, the vast majority of them yielded poor fits or converged to unlikely $J_2$ values.

Using the \verb MERCURY  integrator, we chose a timestep interval ($\sim$1500 seconds for 2001 SN263 and $\sim$4000 seconds for 1994 CC) such that the maximum interpolation error (calculated by comparing interpolated values with those from an integration with one-second timesteps) was at least an order of magnitude less than the smallest observation uncertainty. Cubic spline interpolation was performed to compute state vectors of the satellites at the exact observation epochs to enable comparison with observational data.  For both triple systems, these timesteps were small enough to resolve at least 1/25 of the inner body's orbit. Using the integrator, we were able to obtain the positions and velocities of the satellites with respect to the central body, Alpha, as a function of time. We used line-of-sight vector orientations from the JPL HORIZONS system to project these positions and velocities and to calculate the range and Doppler separations at every observation epoch for each satellite. Comparison between observed and computed range and Doppler values then yielded a measure of the goodness-of-fit according to
\begin{equation}
	\chi^{2} = \displaystyle\sum\limits_{i} \frac{(O_i-C_i)^2}{\sigma_i^2}
\end{equation}
where $\chi^2$ is the chi-square obtained by summing over $i$ comparisons: $O$ and $C$ are the observed and computed values and $\sigma$ is the observation uncertainty.
	
\def\arraystretch{1.4}
\begin{deluxetable}{l r r}
	\centering
	\tablecolumns{3}
	\tablecaption{1994 CC: Range of Initial Conditions \label{gridcc}}
	\startdata
		\hline \hline
		 & Alpha & Maximum Stepsize \\
		\hline
		Mass ($10^{10}$ kg) & 10 - 75 & $\lesssim$25 \\
		$J_2$ & 0.0001 - 0.09 & $\lesssim$0.03 \\
		\hline
		 & Beta (Inner) & Maximum Stepsize \\
		\hline
		Mass ($10^{10}$ kg) & 0.05 - 3 & $\lesssim$1 \\
		$a$ (km) & 0.5 - 4 & $\lesssim$2 \\
		$e$ & 0 - 0.6 & $\lesssim$0.5 \\
		\hline
		& Gamma (Outer) & Maximum Stepsize \\
		\hline
		Mass ($10^{10}$ kg) & 0.0008 - 0.5 & $\lesssim$0.3 \\
		$a$ (km) & 3 - 15 & $\lesssim$5 \\
		$e$ & 0 - 0.7 & $\lesssim$0.5
		\enddata
	\tablenotetext{}{Ranges of initial conditions with the maximum stepsize, defined as the largest interval between two consecutive initial values. Most stepsizes were significantly smaller than those listed. All angles were examined over their full range with typical stepsizes of 5-10 degrees.}
\end{deluxetable}	
	
	Solutions with Gamma more massive than Beta were eliminated because they are inconsistent with radar size estimates. In our list of best fits, for 2001 SN263 our lowest reduced chi-square was $\sim$0.093 with 496 degrees of freedom (DOF), and in the case of 1994 CC, our lowest reduced chi-square was $\sim$0.27 with 364 degrees of freedom. The likely cause of a reduced chi-square less than 1 is the overestimation of observation uncertainties. We considered only those solutions within a 1-$\sigma$ increase in the lowest reduced chi-square value (Press et al. 1992):
	\begin{equation}
		\chi_{\nu,1\sigma}^{2} = \chi_{\nu,\rm min}^{2} + \chi_{\nu,\rm min}^{2}*\dfrac{\sqrt{2*\rm DOF}}{\rm DOF}
	\end{equation}
where $\chi_{\nu,1\sigma}^{2}$ represents the chi-square value calculated from a 1-$\sigma$ increase of the minimum chi-square value $\chi_{\nu,\rm min}^{2}$. This resulted in $\chi_{\nu,1\sigma}^{2} \sim$ 0.1 for 2001 SN263 and $\chi_{\nu,1\sigma}^{2} \sim$ 0.29 for 1994 CC. Such criteria further narrowed our list of fits, but contained solutions that were near-duplicates. For ease of presentation we marked and removed nearly duplicate fits that met both of the following conservative criteria: differences in masses (Alpha, Beta, and Gamma) were less than 2\% and differences in orbital orientations were less than 5 degrees. The other orbital elements were relatively consistent, and we did not incorporate them into the filter.	
	
\subsection{Long-Term Integrations: Stability}

	To further discriminate between our remaining least-squares solutions and to study stability, we looked at their orbital evolution over time through long-term integrations. Only orbital solutions that were stable over the course of longer-term numerical integrations were regarded as satisfactory solutions. Since the dynamical lifetimes of these asteroids are $\sim$10 Myrs (Gladman et al. 1997), we ran extensive stability tests ($\sim$5 Myrs) for our remaining orbital fits. This resulted in a subset of likely solutions that met our constraints. The timestep interval was chosen such that we could resolve at least 1/20 of the inner body's orbit for both triple systems, which is small enough to capture the dynamics of the system. In these long-term integrations, we incorporated collisions and ejections using a hybrid symplectic/Bulirsch-Stoer algorithm. 
	
	To include collisions, it was necessary to specify physical sizes for all components. For Alpha, we used their known radii: for 2001 SN263, R $\sim$ 1.3 km (Nolan et al. 2008b) and for 1994 CC, R $\sim$ 315 m (Brozovic et al. 2010). For Beta and Gamma we used our mass solutions and scaled from Alpha assuming identical bulk densities.
	
	Ejections were defined as events where the satellite was no longer within 1 Hill radius of Alpha. The Hill radius equation for Alpha is the following:
\begin{equation}
	r_{\rm Hill} = a_{\odot}\left(\dfrac{M_{\rm Alpha}}{3M_{\Sun}}\right)^{1/3}
\end{equation}
	where $M_{\rm Alpha}$ and $M_{\Sun}$ represent the masses of Alpha and the Sun, respectively, and $a_{\odot}$ is Alpha's semi-major axis with respect to the Sun.
	
\subsection{Precession Rates}

Non-Keplerian effects potentially provide powerful constraints on component masses. We measured the precession rates of our best-fit solutions and compared them to analytical expressions of the precession due to an oblate primary and secular perturbations. For primary oblateness described by $J_2$, the analytical expressions are (Murray \& Dermott 1999):
\begin{equation}
	\dfrac{d\omega}{dt} \approx \dfrac{3}{2}\dfrac{nJ_2}{(1-e^2)^2}\left(\dfrac{R}{a}\right)^2\left(\dfrac{5}{2}\cos^2I-\dfrac{1}{2}\right)  \vspace{2 mm}
\end{equation}
\begin{equation}
	\dfrac{d\Omega}{dt} \approx -\dfrac{3}{2}nJ_2\left(\dfrac{R}{a}\right)^2\dfrac{\cos I}{(1-e^2)^2}
\end{equation}
where $\omega$ and $\Omega$ represent the argument of pericenter and the longitude of the ascending node, respectively. The precession rate for the longitude of pericenter is simply the sum of the rates for the argument of pericenter and longitude of the ascending node. The other variables are defined as follows: $n$ is the mean motion of the satellite, $J_2$ is the unitless quantity describing the oblateness of the central body, $e$ is the eccentricity, $R$ is the radius of the central body in the same units as the satellite's semi-major axis $a$, and $I$ is the orbital inclination with respect to Alpha's equator.

	We also calculated the analytical precession rates due to secular perturbations. Over long timescales and in the absence of any mean-motion resonances or additional perturbations, the secular equations adequately describe the evolution of orbital elements. The secular equations can be obtained by examining the solution to Lagrange's equations corresponding to the secular part (terms that do not depend on the mean longitudes of either body) of the disturbing function, which describes a body's gravitational perturbations by other bodies (Murray \& Dermott 1999). The secular equations can be written with convenient variables:
\begin{equation}
	h_j = e_j\sin(\varpi_j) \hspace{20 mm} \vspace{1 mm} p_j = I_j\sin(\Omega_j)
\end{equation}
\begin{equation}
	k_j = e_j\cos(\varpi_j) \hspace{20 mm} \vspace{1 mm} q_j = I_j\cos(\Omega_j)
\end{equation}
where $h$, $k$, $p$, and $q$ are new variables as defined above, $e$ is the eccentricity, $\varpi$ is the longitude of pericenter, $I$ is the inclination with respect to Alpha's equator, $\Omega$ is the longitude of the ascending node, and $j$ denotes the body being perturbed. Their solutions are of the following form:
\begin{equation}
	h_j = \displaystyle\sum\limits_{i=1}^2e_{ji}\sin(g_it+\beta_i) \vspace{1 mm} \hspace{5 mm}
	p_j = \displaystyle\sum\limits_{i=1}^2I_{ji}\sin(f_it+\gamma_i) \vspace{1 mm}
\end{equation}
\begin{equation}
	k_j = \displaystyle\sum\limits_{i=1}^2e_{ji}\cos(g_it+\beta_i) \vspace{1 mm} \hspace{5 mm}
	q_j = \displaystyle\sum\limits_{i=1}^2I_{ji}\cos(f_it+\gamma_i) \vspace{1 mm}
\end{equation}
where $i$ denotes the eigenmodes, $g_i$ and $f_i$ represent the eigenfrequencies, and $\beta_i$ and $\gamma_i$ are the phases (Murray \& Dermott 1999). From these secular solutions, we were able to calculate $e(t)$, $\varpi(t)$, $I(t)$, and $\Omega(t)$ to find the analytical precession rates corresponding to secular theory.

Our measurements of the precession rates used our best-fit solutions to the observations. We numerically integrated these solutions to detect the evolution of the argument of pericenter, longitude of ascending node, and longitude of pericenter for both satellites in each system. Due to fast, short-term fluctuations in the orbital elements as a function of time, it was necessary to perform linear regressions to obtain estimates of the rates over typical timescales of 100s of days. In the case of 2001 SN263 Beta, we looked at longer timescales of thousands of days due to its relatively slow precession. Since the rates also changed over time, we computed them near the epoch at which we solved for the orbits. By measuring the numerical rates at which these orbital elements precess, we were able to compare them with those calculated from analytical methods above, which included $J_2$ and secular contributions.

	
\subsection{Resonances}

We searched for mean-motion resonances between Beta and Gamma in both systems. To do this, we first looked for integer values (-150 to +150) of $j_1$ and $j_2$ where the following resonant argument varied ``slowly" ($<$100 deg/day):
\begin{equation}
	\phi \approx j_1 n_2 + j_2 n_1
\end{equation}
	where $n_1$ and $n_2$ denote the mean motions of the inner and outer bodies, respectively. For values of $j_1$ and $j_2$ that met our first criterion, we searched through integer values (-30 to +30) of $j_3$, $j_4$, $j_5$, and $j_6$ for which there was libration of the general resonant argument (Murray \& Dermott 1999):
\begin{equation}
	\phi = j_1 \lambda_2 + j_2 \lambda_1 + j_3 \varpi_2 + j_4 \varpi_1 + j_5 \Omega_2 + j_6 \Omega_1
\end{equation}
where $\lambda$ is the mean longitude, $\varpi$ is the longitude of pericenter, and $\Omega$ is the longitude of the ascending node. The d'Alembert relation:
\begin{equation}
 \displaystyle\sum\limits_{i=1}^6j_i = 0
\end{equation}
describes how the integer values of $j_i$ $(1 \leq i \leq 6)$ are related. Lastly, we repeated the second criterion for different permutations of the satellites' mean longitudes representing the highest and lowest possible values that captured the full range of uncertainties on mean motions.

\section{Orbit-Fitting Results}

	In this section, we describe the best-fit orbital solutions that passed the constraints described in Table \ref{constraints} and Section 2. Our initial fits resulted in 174 and 901 possible solutions within a 1-$\sigma$ increase in the lowest reduced chi-square value for 2001 SN263 and 1994 CC, respectively. Of these, 53 and 582 were eliminated by the mass bounds, $J_2$ constraints, and duplicate filtering.  After further eliminating solutions that did not meet our long-term integration criterion, we were left with a list of stable solutions for both systems (2001 SN263: 113 fits, 1994 CC: 262 fits), which were used for post-orbit-fitting determinations of orbital parameter uncertainties. The majority of our unstable fits reached instability quickly--typically within 1000 years for both systems.
	
	A significant fraction (2001 SN263: $\sim$25\%; 1994 CC: $\sim$45\%) of our fits resulted in a retrograde orbit of either Beta or Gamma (orbital direction opposite to Alpha's spin direction). This occurred because we were able to fit the data with a wide range of spin axis orientations for Alpha even though the orbital orientations of Beta and Gamma were fairly well determined. While it is more likely that the orbits of Beta and Gamma are prograde with respect to Alpha, some of our retrograde solutions are stable over $\sim$Myrs (in the case of no external perturber) and at this point we cannot rule them out.
	
\begin{figure}[h!]
	\centering
	\includegraphics[scale=0.35]{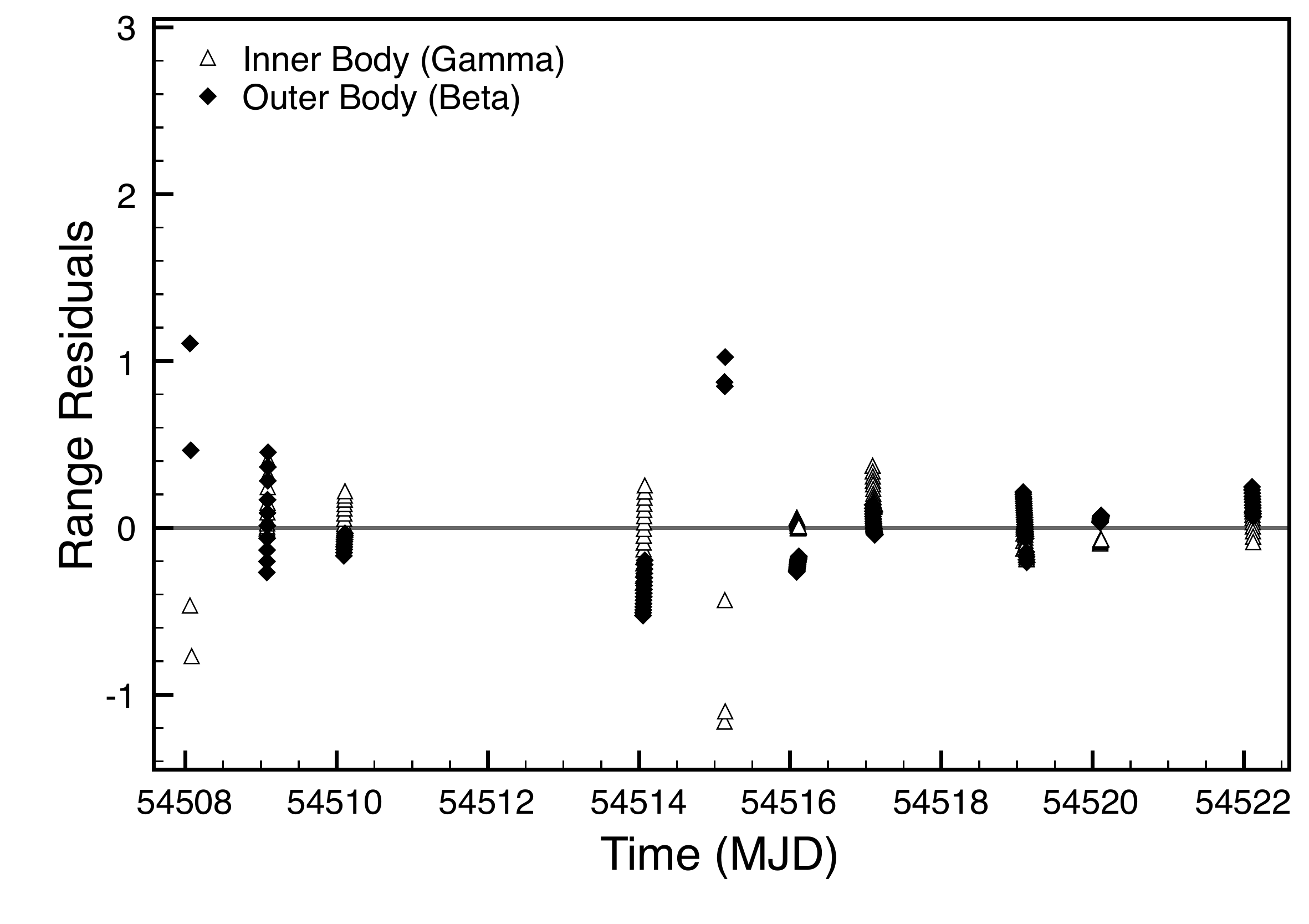}
	\includegraphics[scale=0.35]{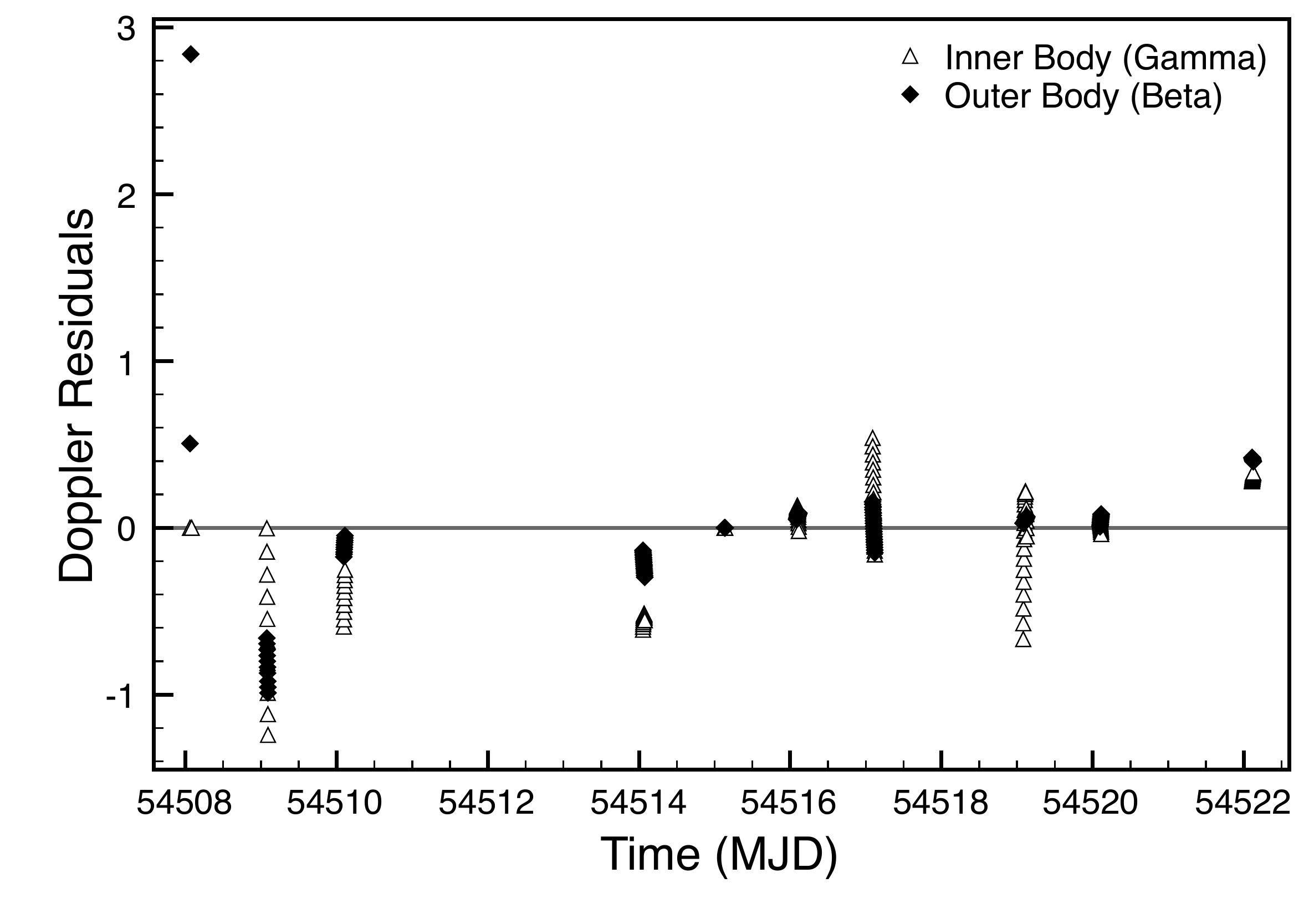}
	\caption{2001 SN263 Best-Fit Solution: Range and Doppler Residuals. The second Doppler residual at MJD 54508.0708 is a clear outlier. We obtained similar solutions whether we included this data point or not. We also note that there seems to be an apparent sinusoidal signature in the residuals for 2001 SN263, which may indicate that our model is not capturing the full dynamics of the system. Other good orbital solutions with plausible mass ratios and $J_2$ values also produce similar residuals.} \label{snres}
\end{figure}

\begin{figure}[tbh]
		\centering
		\includegraphics[scale=0.35]{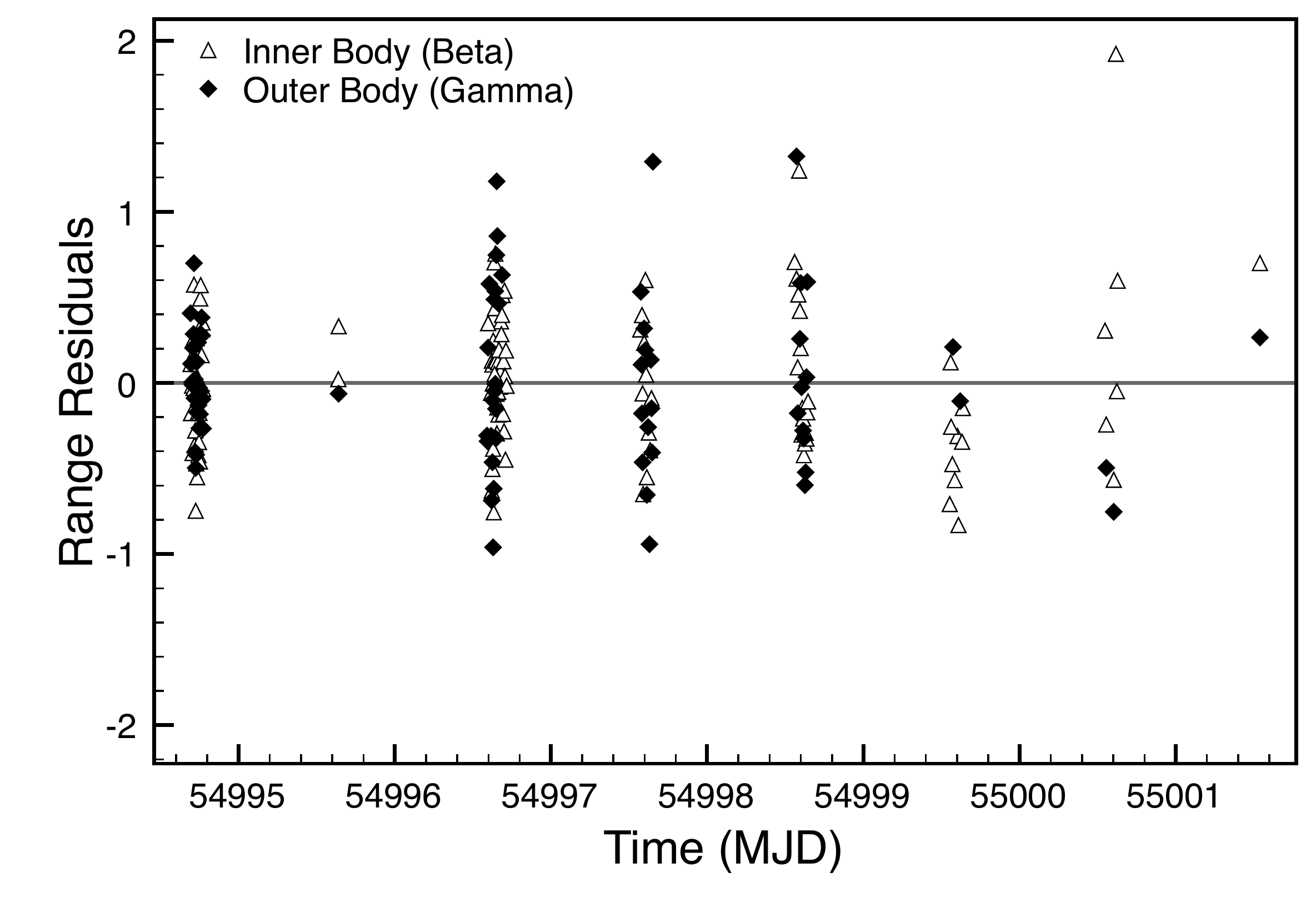}
		\includegraphics[scale=0.35]{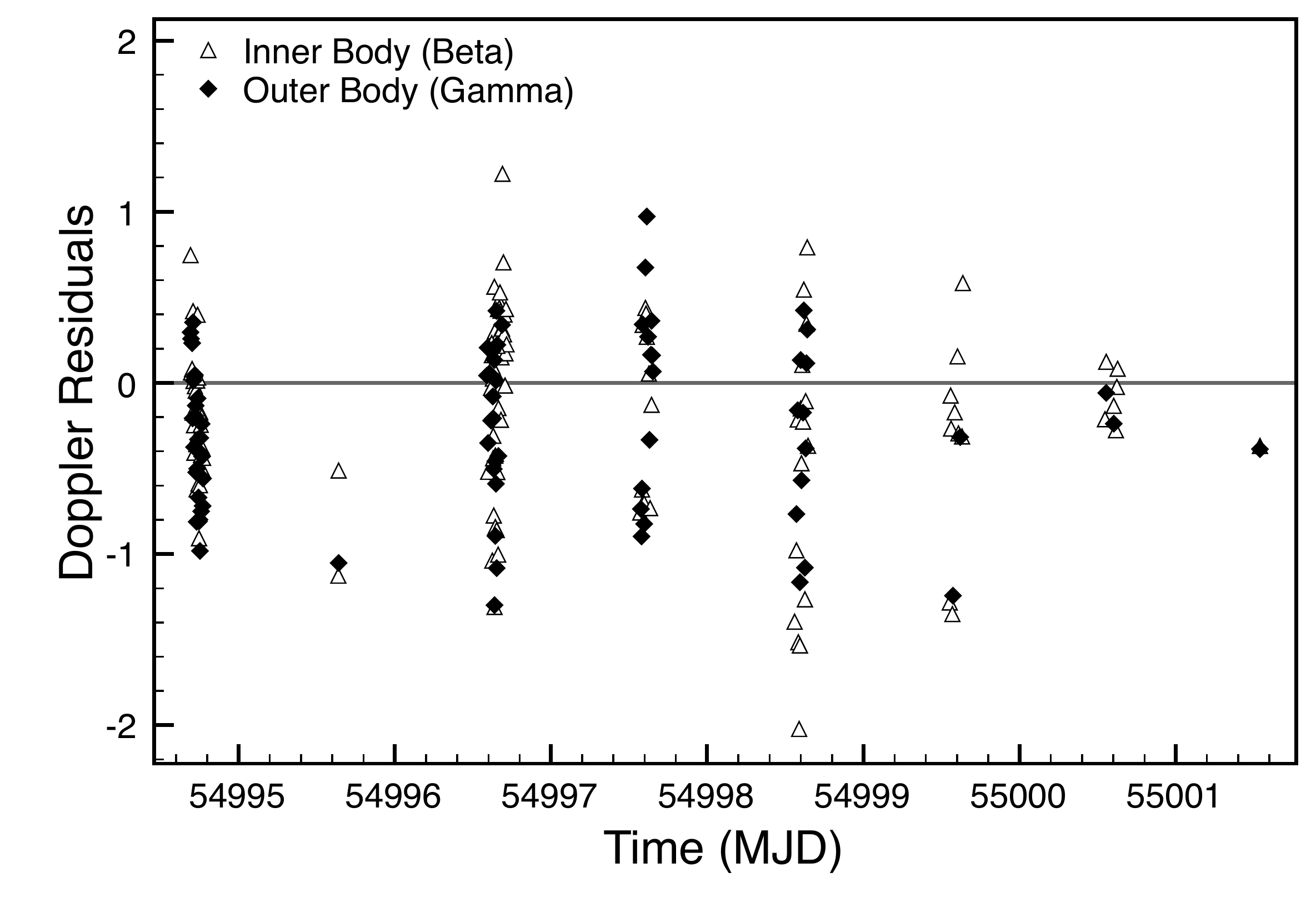}
		\caption{1994 CC Best-Fit Solution: Range and Doppler Residuals.} \label{ccres}
\end{figure}

	Diagrams showing the best-fit orbits of the satellites in 2001 SN263 and 1994 CC projected onto Alpha's equatorial plane are shown in Figure \ref{diagrams}. The best-fit models match the data well (Figures \ref{snres} and \ref{ccres}), where the residuals shown are defined as (observation - model)/uncertainty. For both triples, the total angular momentum budget (orbital and spin) is dominated by Alpha's spin angular momentum. For 2001 SN263, Alpha contributes 77\%, Gamma (inner body) contributes 4\%, and Beta (outer body) contributes 19\% of the total angular momentum. For 1994 CC, Alpha contributes 85\%, Beta (inner body) contributes 12\%, and Gamma (outer body) contributes 3\% of the total angular momentum. In both systems, $\sim$97\% of the angular momentum is contained in the spin of Alpha and the orbit of Beta, which justifies a posteriori our decision to align Beta's orbit with Alpha's equatorial plane, consistent with a spin-up formation process.

	\def\arraystretch{1.4}
	\begin{deluxetable}{l r r}
		\centering
		\tablecolumns{3}
		\tablecaption{2001 SN263: Best-Fit Parameters and Formal 1-$\sigma$ Errors\label{bestsn}}
		\startdata
			\hline \hline
			 & Gamma (inner) & Beta (outer) \\
			\hline
			Mass ($10^{10}$ kg) & 9.773 $\pm$ 3.273 & 24.039 $\pm$ 7.531 \\
			$a$ (km) & 3.804 $\pm$ 0.002 & 16.633 $\pm$ 0.163 \\
			$e$ & 0.016 $\pm$ 0.002 & 0.015 $\pm$ 0.009 \\
			$i$ (deg) & 165.045 $\pm$ 12.409 & 157.486 $\pm$ 1.819 \\
			$\omega$ (deg) & 292.435 $\pm$ 53.481 & 131.249 $\pm$ 21.918  \\
			$\Omega$ (deg) & 198.689 $\pm$ 61.292  & 161.144 $\pm$ 13.055 \\
			$M$ (deg) & 248.816 $\pm$ 11.509 & 212.658 $\pm$ 10.691  \\
			$P$ (days) & 0.686 $\pm$ 0.00159 & 6.225 $\pm$ 0.0953 \\
			\hline \hline
			 & \multicolumn{2}{r}{Alpha (central body)} \\
			\hline
			Mass ($10^{10}$ kg) & \multicolumn{2}{r}{917.466 $\pm$ 2.235}  \\
			$J_2$ & \multicolumn{2}{r}{0.013 $\pm$ 0.008} \\
			Pole Solution (deg) & \multicolumn{2}{r}{RA: 71.144 $\pm$ 13.055} \\
			& \multicolumn{2}{r}{DEC: -67.486 $\pm$ 1.819}
			\enddata
		\tablenotetext{}{The masses are listed in $10^{10}$ kg, $a$ is the semi-major axis, $e$ is the eccentricity, $i$ is the inclination, $\omega$ is the argument of pericenter, $\Omega$ is the longitude of the ascending node, $M$ is the mean anomaly at epoch, and $P$ is the period. These orbital elements are valid at MJD 54509 in the equatorial frame of J2000. Alpha's pole solution is given in right ascension (RA) and declination (DEC). This table lists formal 1-$\sigma$ statistical errors; see text for adopted 1-$\sigma$ uncertainties.}
	\end{deluxetable}	

\subsection{2001 SN263} 
	
	The best-fit parameters (Table \ref{bestsn}) are valid at the epoch MJD 54509.0, where the reduced chi-square for this orbital solution is $\chi_{\nu}^{2}$ = 0.099 (DOF = 496). While a sizeable fraction of our orbital fits had a lower chi-square, the adopted solution described in this section and Table \ref{bestsn} has the most plausible combination of Beta/Gamma mass ratio and $J_2$ value (lower chi-square solutions with a Beta/Gamma mass ratio less than 2 or a $J_2$ of 0 were not considered based off of radar size and shape estimates). For discussion regarding masses and $J_2$, see Section 3.3. The formal 1-$\sigma$ errors listed in Table \ref{bestsn} certainly underestimate the actual errors; here we list plausible 1-$\sigma$ uncertainties alongside parameter values with guard digits by examining the range of parameter values in our 113 acceptable fits. The mass of Alpha is $917.466_{-5}^{+19} \times 10^{10}$ kg, and the masses of Gamma and Beta are $9.773 \pm 7 \times 10^{10}$ and $24.039_{-17}^{+7} \times 10^{10}$ kg, respectively. Using preliminary size estimates from radar images, we calculate a density of $\sim$0.997 g/cm$^3$ for Alpha. If we apply Alpha's density to the range of satellite masses that are within uncertainties, their equivalent radii range from 188 - 342 m for Gamma and 213 - 420 m for Beta.

	The orbit of Gamma has a semi-major axis of $3.804_{-0.02}^{+0.01}$ km and an eccentricity of $0.016_{-0}^{+0.005}$. The orbit of Beta is also nearly circular with an eccentricity of $0.015_{-0.010}^{+0.022}$ and a semi-major axis of $16.633_{-0.38}^{+0.39}$ km. The orbital periods of Gamma and Beta are $0.686 \pm 0.01$ and $6.225 \pm 0.5$ days, respectively. Assuming that Beta orbits in Alpha's equatorial plane, the mutual inclination between the satellites' orbital planes is $\sim$14 degrees. We estimate orbit pole angular uncertainties of $\sim$10 degrees for Beta and $\sim$15 degrees for Gamma. Alpha's floating $J_2$ value converged to $0.013_{-0.013}^{+0.050}$. 

\begin{figure*}[htbp]
	\centering
	\mbox{\subfigure{\includegraphics[scale=0.5]{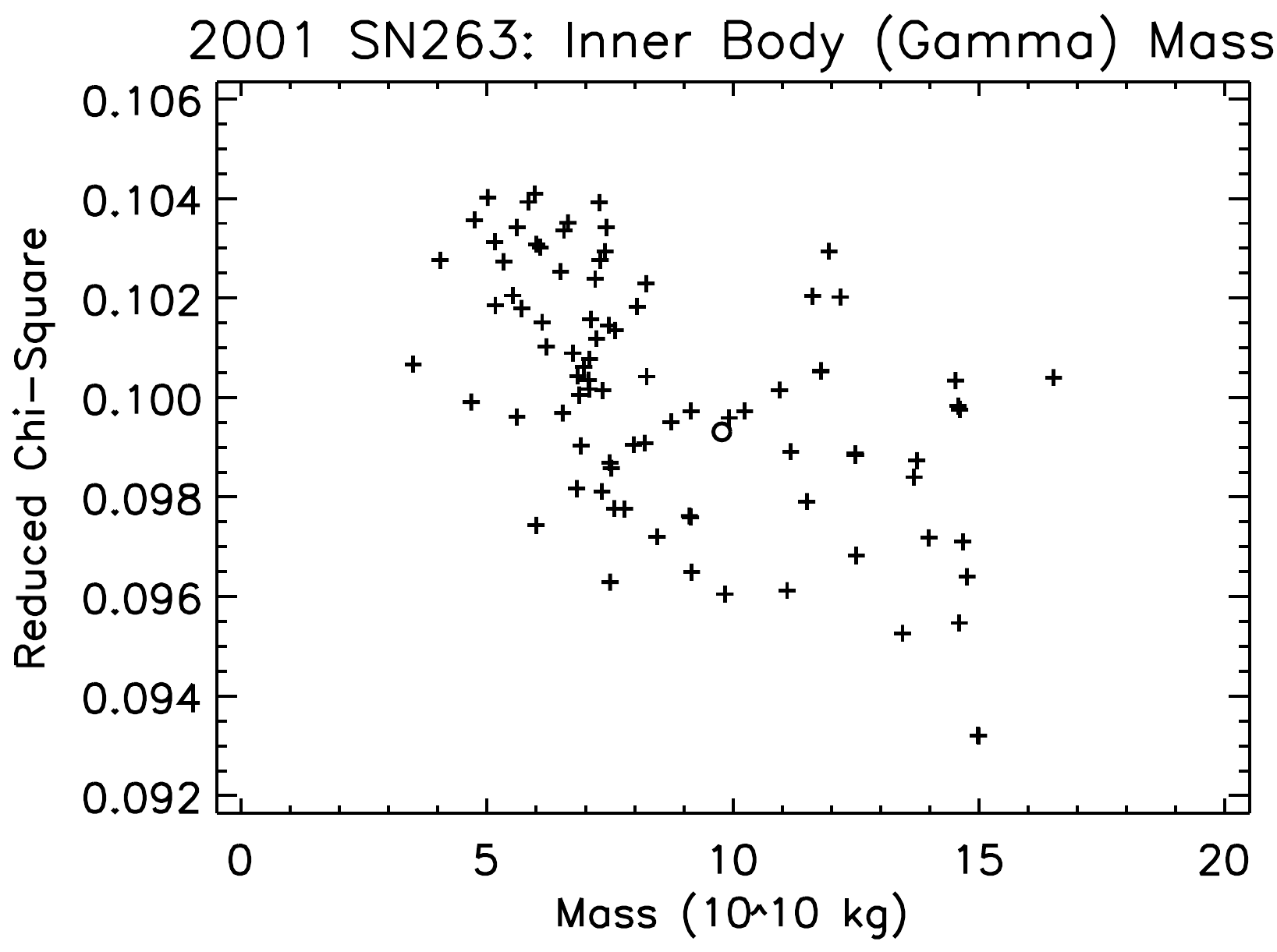}}\quad
	\subfigure{\includegraphics[scale=0.5]{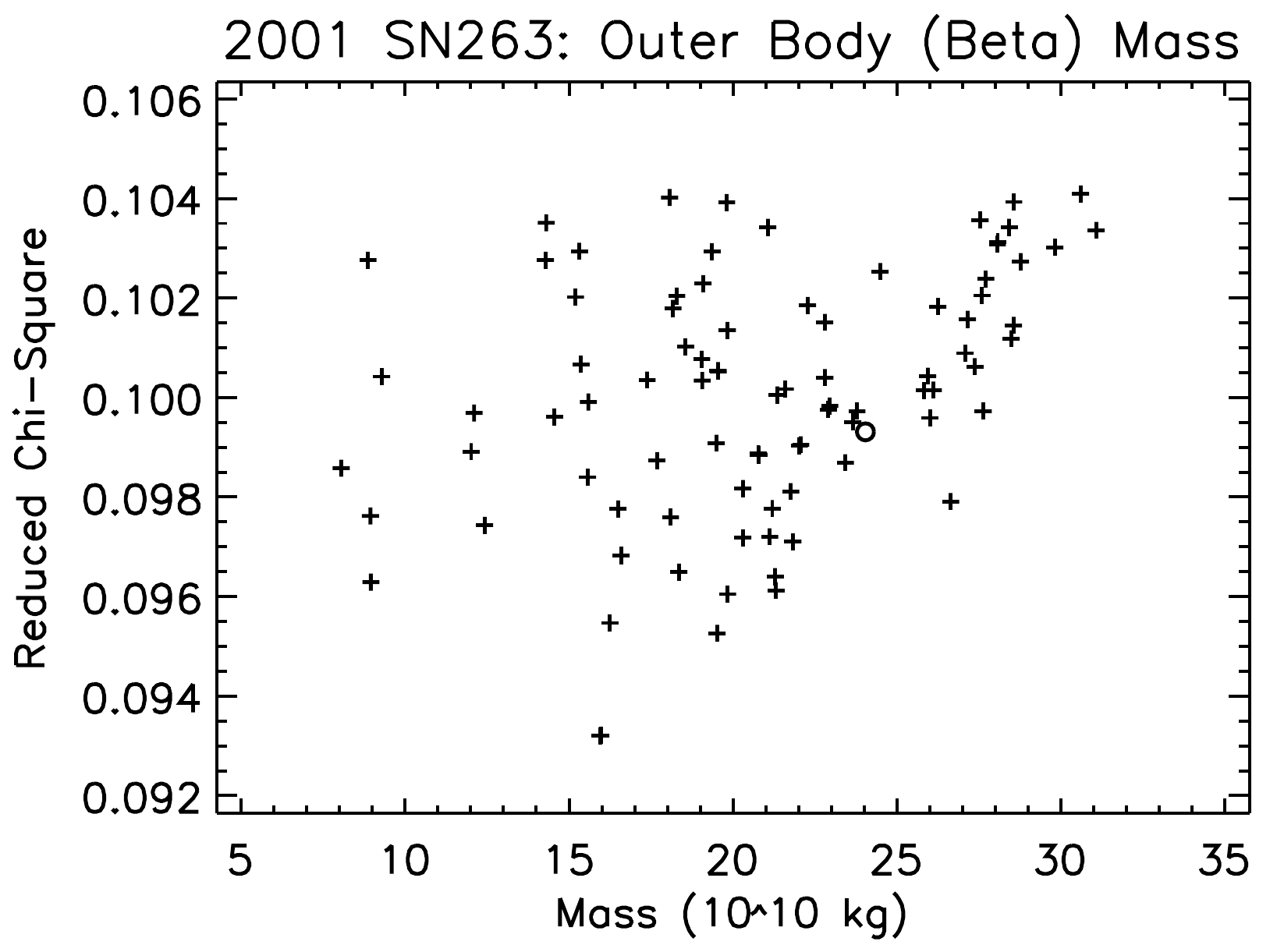}}}
	\mbox{\subfigure{\includegraphics[scale=0.5]{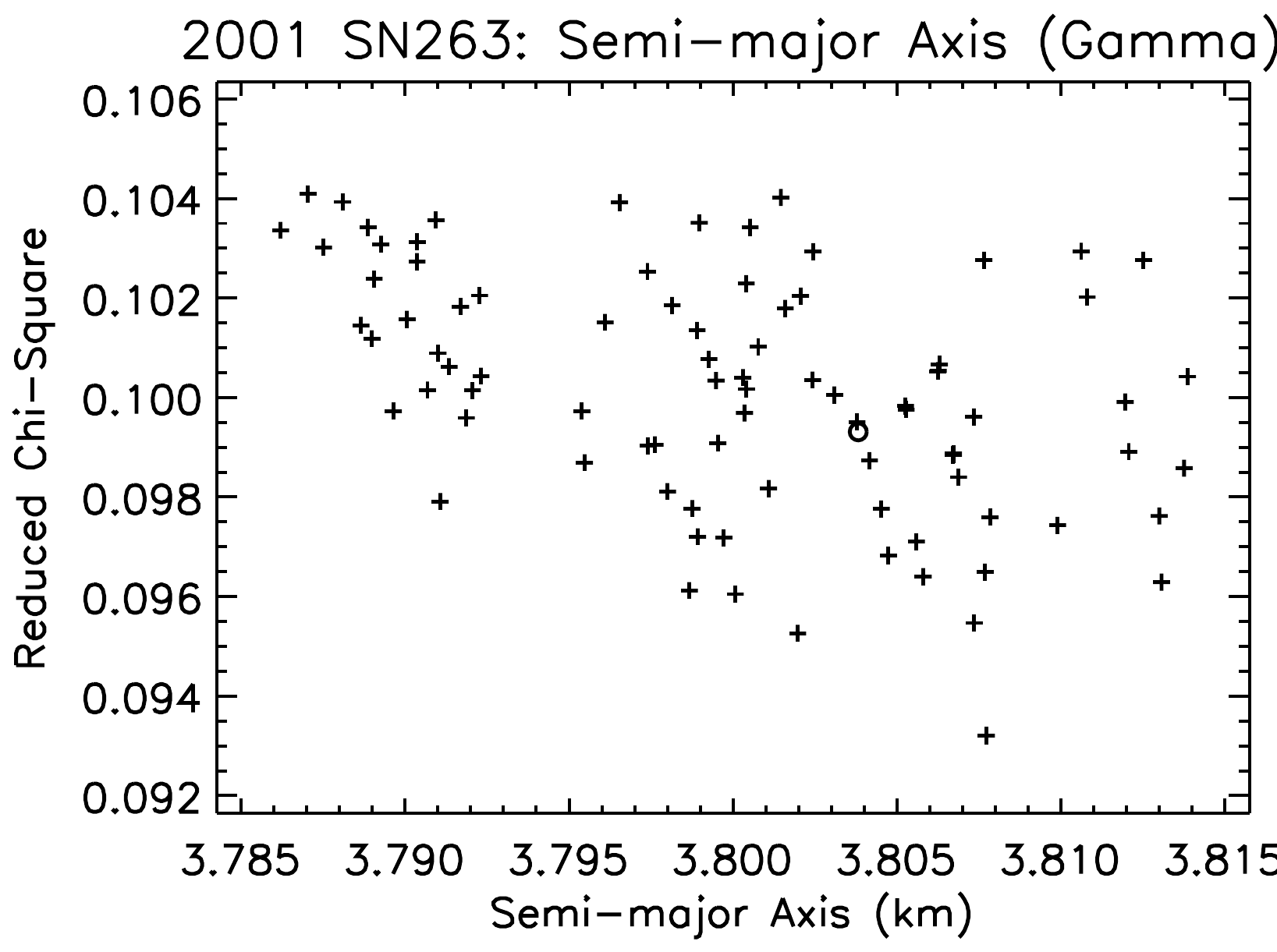}}\quad
	\subfigure{\includegraphics[scale=0.5]{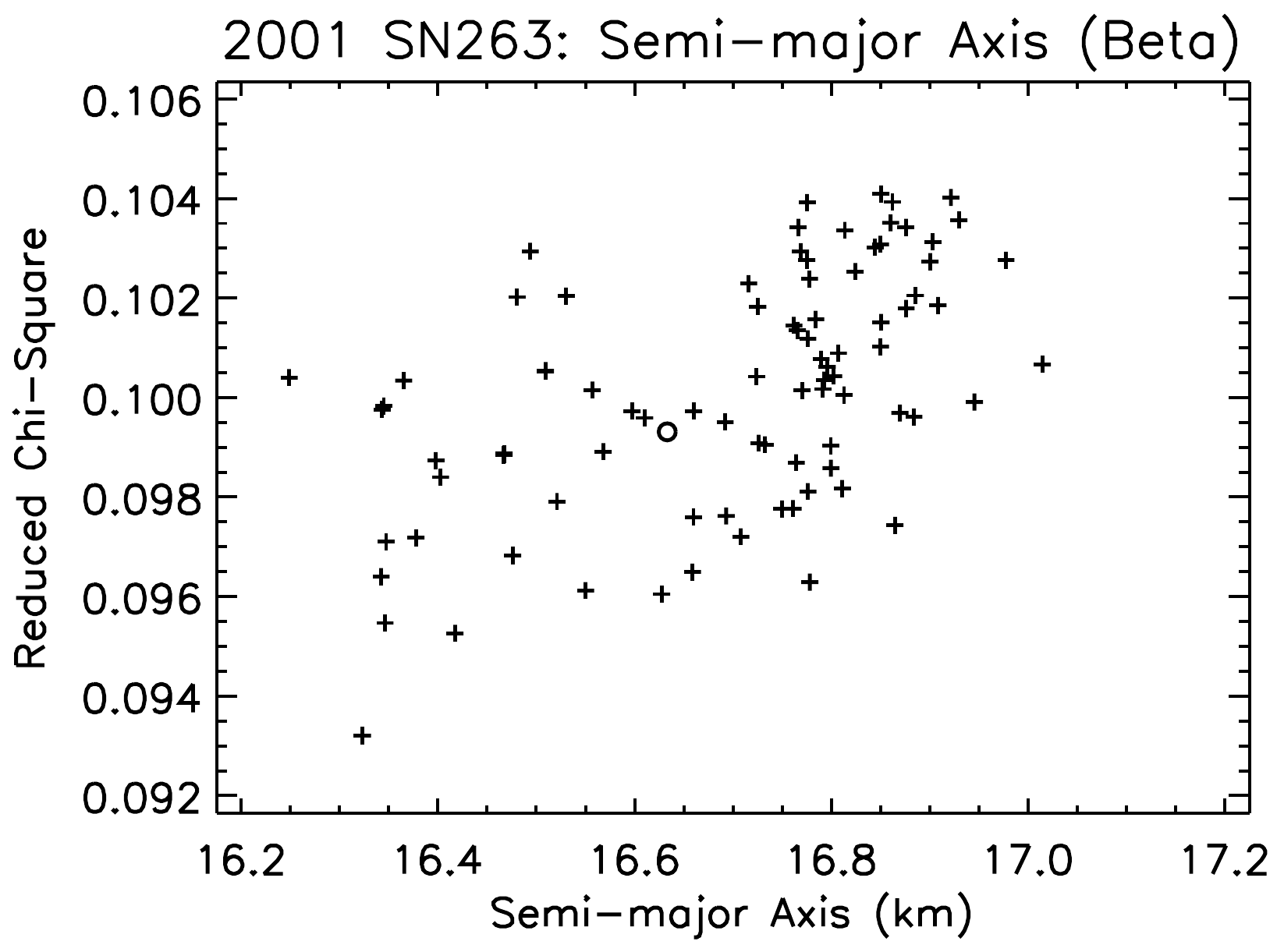}}}
	\mbox{\subfigure{\includegraphics[scale=0.5]{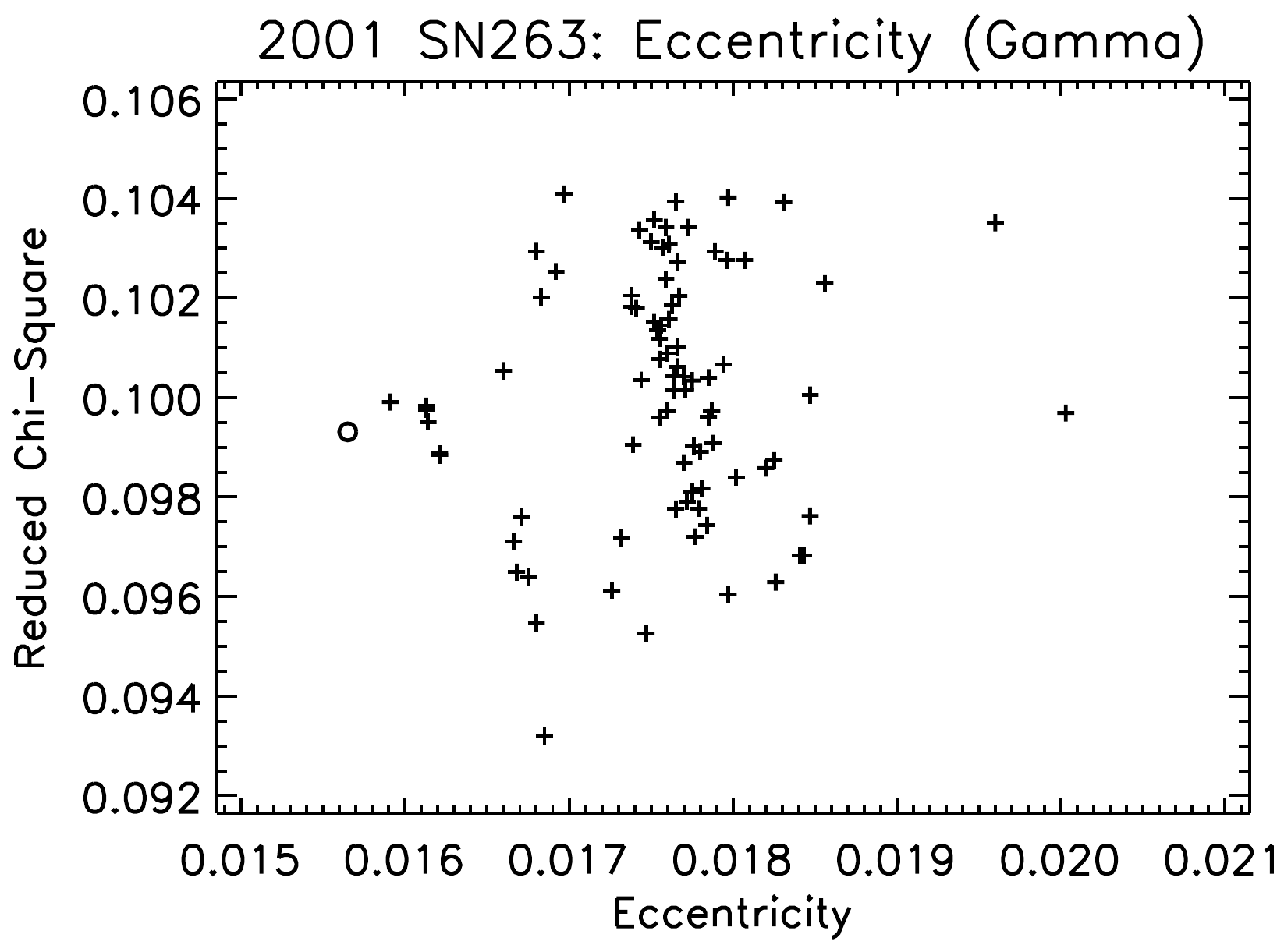}}\quad
	\subfigure{\includegraphics[scale=0.5]{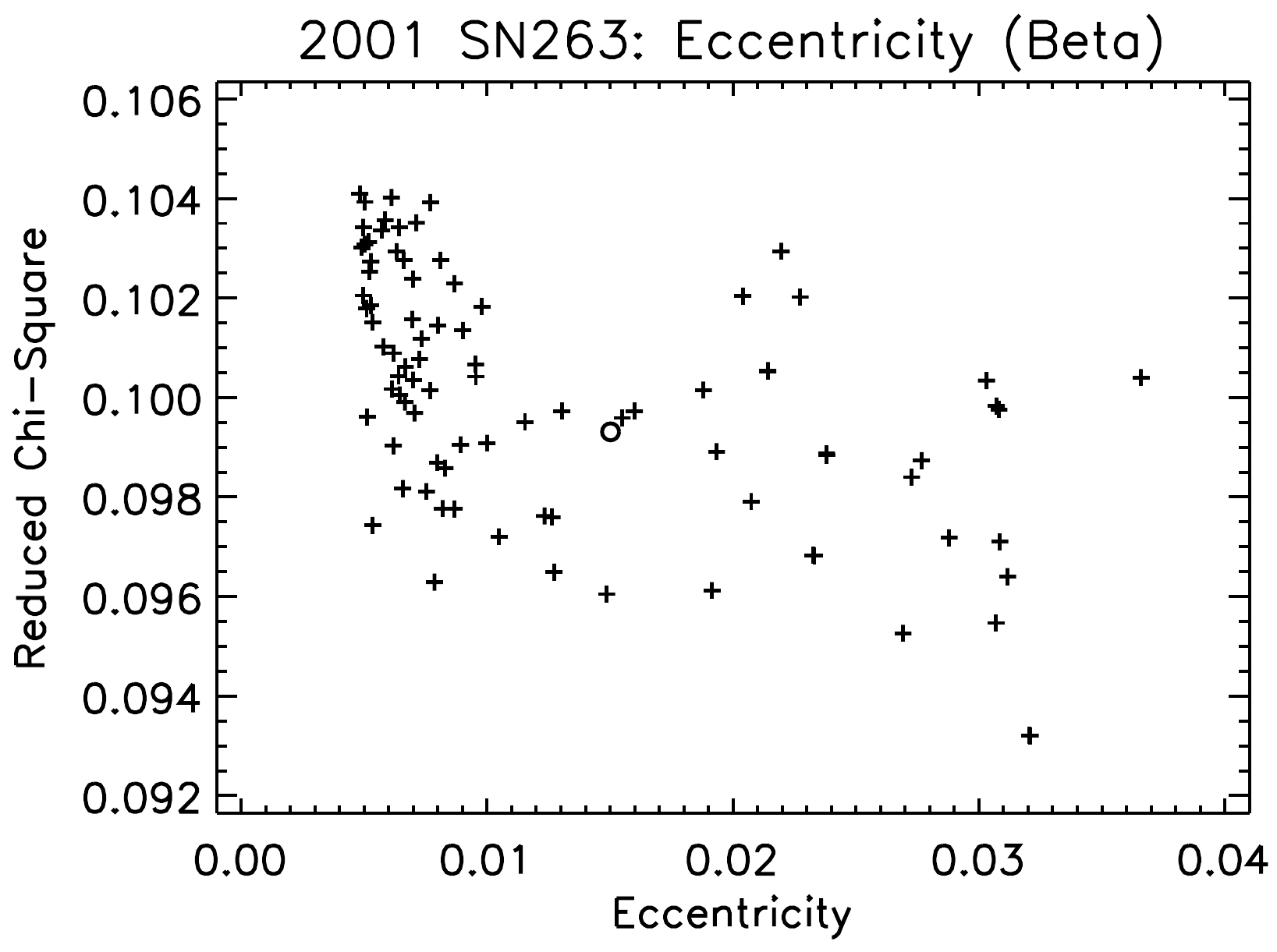}}}
	\caption{2001 SN263: Reduced chi-square as a function of orbital parameters: mass, semi-major axis, and eccentricity. The points plotted here include the statistically equivalent best-fit orbital solutions that passed the constraints listed in Table \ref{constraints} and Section 2.  Our adopted value is shown as an open circle, and solutions with lower chi-squares were ruled out on the basis of implausible $J_2$ and Beta/Gamma mass ratios.} \label{snchi1}
\end{figure*}

\def\arraystretch{1.4}
\begin{deluxetable*}{l r r r r r r r r r}
	\tablecaption{2001 SN263 Precession Rates \label{snrates}} \\
	\startdata
	\hline \hline
	 & {\bf Gamma (inner)} & & & & {\bf Beta (outer)} & & & \\
	 & Analytical: & & & Numerical  & Analytical: & & & Numerical  \\
	 & Secular & $J_2$ & Total & Total  & Secular & $J_2$ & Total & Total  \\
	\hline
	$\dot{\omega}$ (deg/day) & 0.314 & 2.239 & 2.552 & 2.466  & 0.110 & 0.014 & 0.124 & 0.093 \\
	$\dot{\Omega}$ (deg/day) & -0.167 & -1.171 & -1.338 & -1.295 & -0.081 & -0.007 & -0.087 & -0.063\\
	$\dot{\varpi}$ (deg/day) & 0.146 & 1.068 & 1.214 & 1.120  & 0.030 & 0.007 & 0.037 & 0.030
	\enddata
	\tablenotetext{}{This table includes the argument of pericenter rate $\dot{\omega}$, the longitude of the ascending node rate $\dot{\Omega}$, and the longitude of pericenter rate $\dot{\varpi}$. For each satellite, the first 3 columns represent our analytical calculations and the fourth column shows our measured numerical rates.}
\end{deluxetable*}

\begin{figure*}[htbp]
	\centering
	\mbox{\subfigure{\includegraphics[scale=0.5]{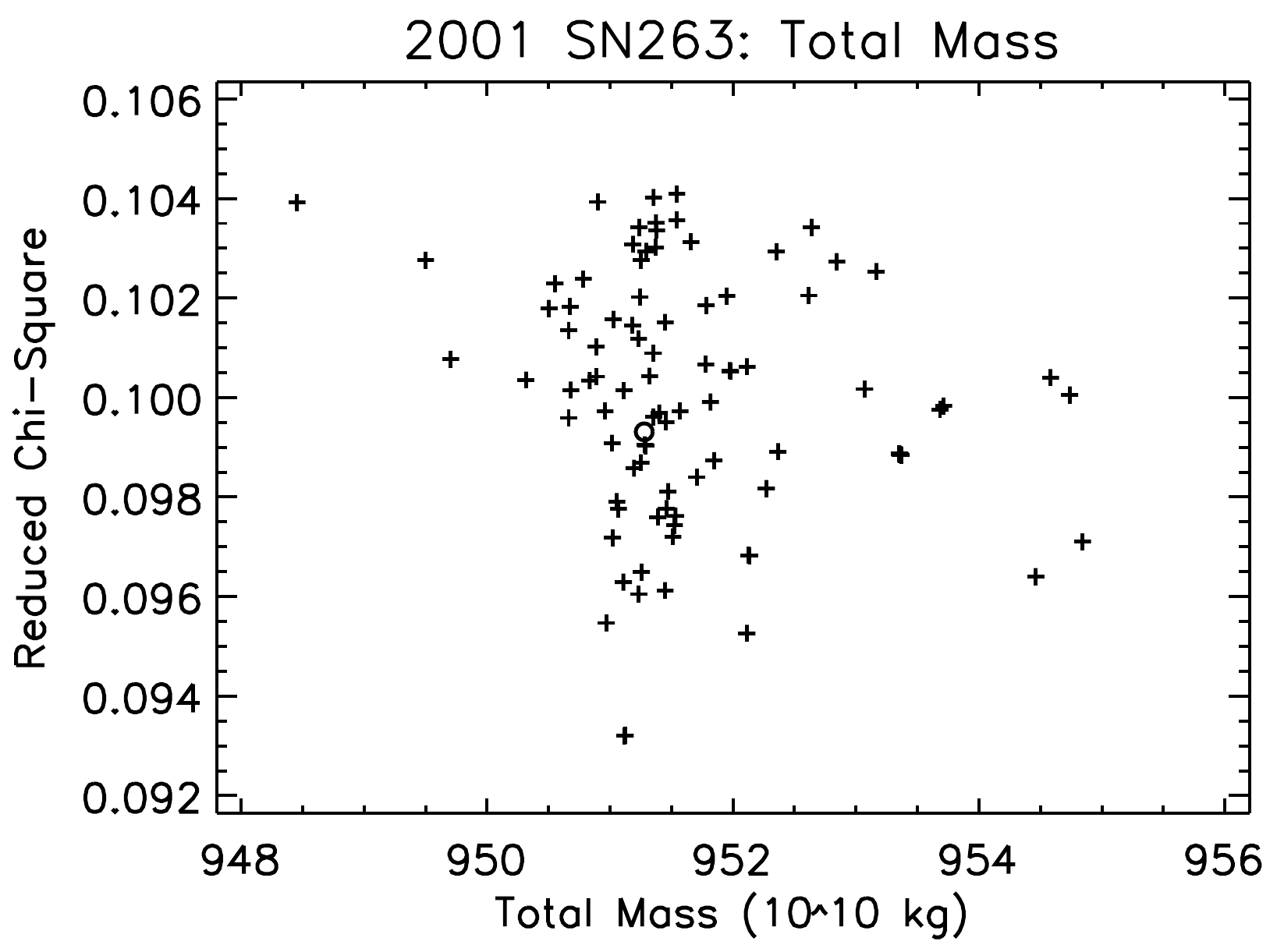}}\quad
	\subfigure{\includegraphics[scale=0.5]{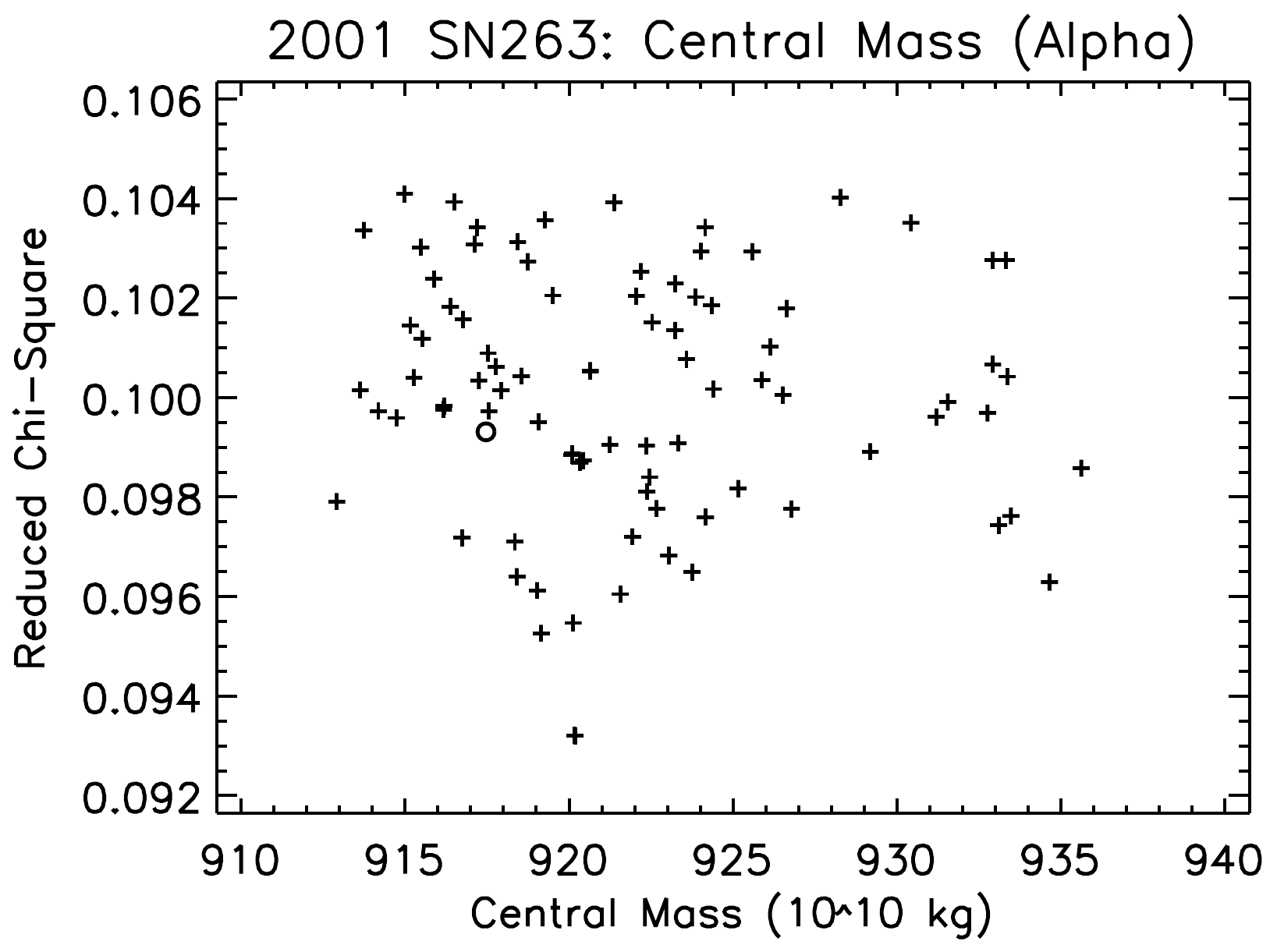}}}
	\mbox{\subfigure{\includegraphics[scale=0.5]{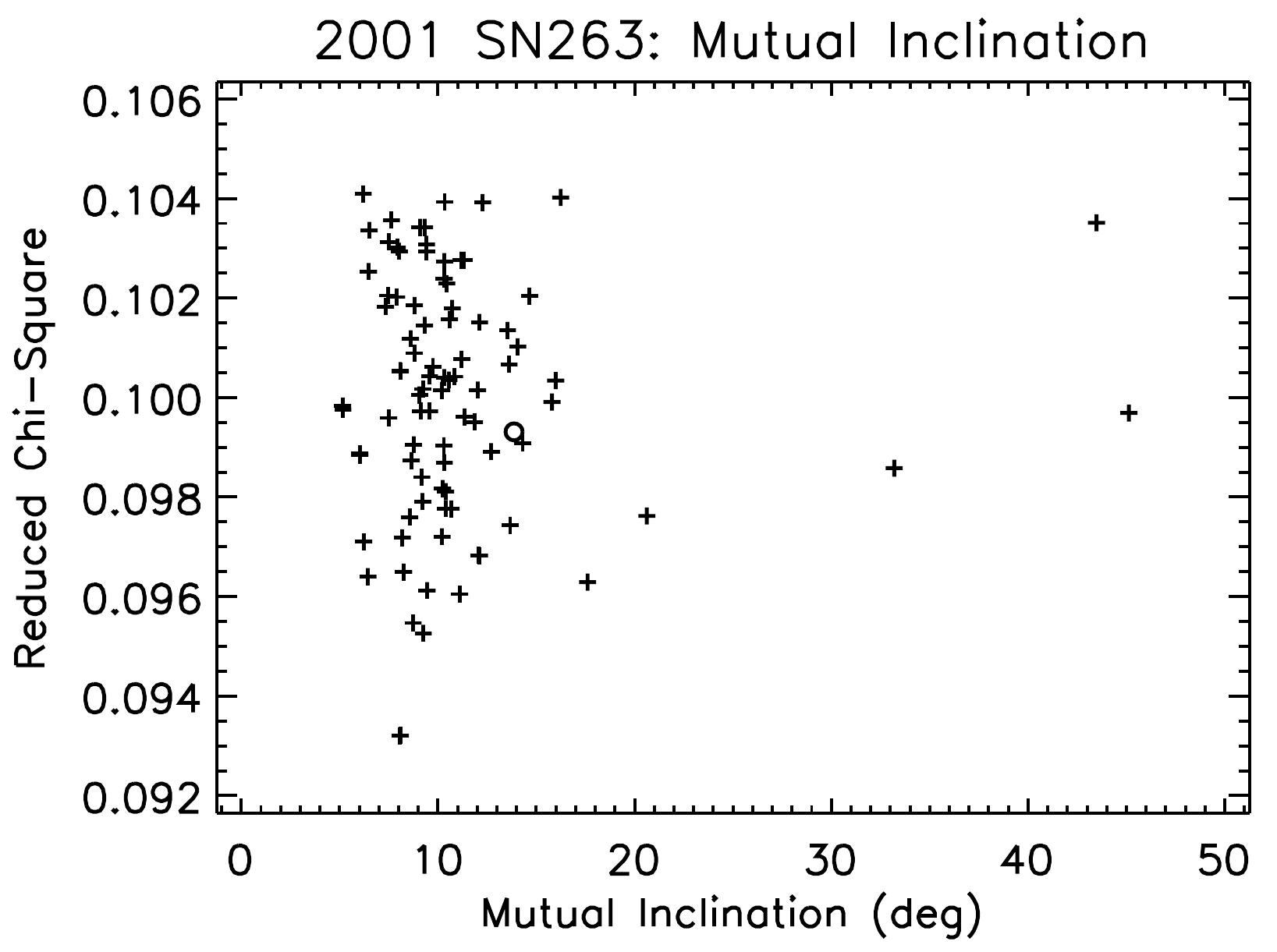}}\quad
	\subfigure{\includegraphics[scale=0.5]{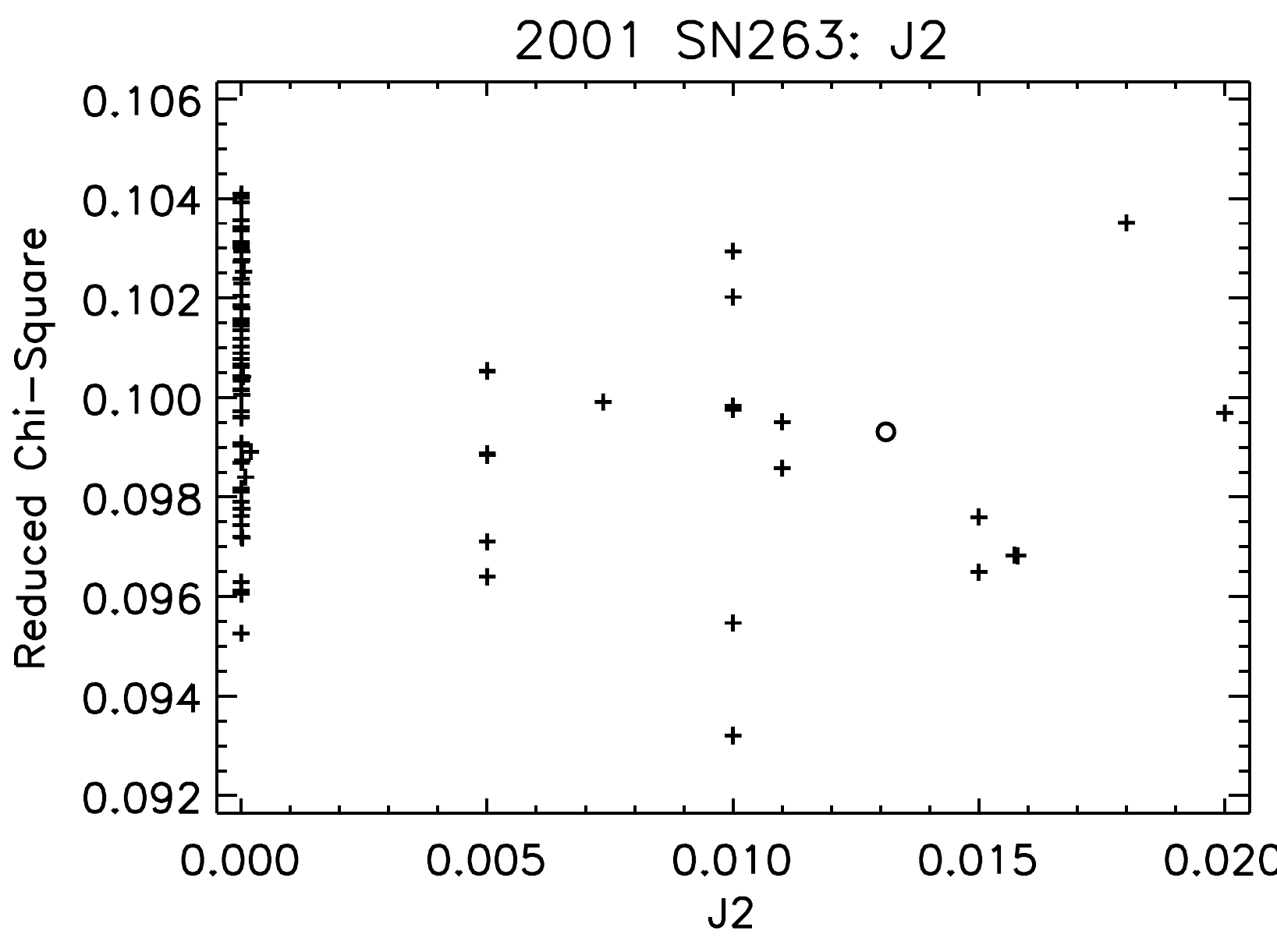}}}
	\caption{2001 SN263: Reduced chi-square as a function of masses, mutual inclination, and $J_2$. The points plotted here include the statistically equivalent best-fit orbital solutions that passed the constraints listed in Table \ref{constraints} and Section 2. Our adopted value is shown as an open circle, and solutions with lower chi-squares were ruled out on the basis of implausible $J_2$ and Beta/Gamma mass ratios.} \label{snchi2}
\end{figure*}
\clearpage

Plots showing our statistically equivalent (within 1-$\sigma$ level) best fits are shown in Figures \ref{snchi1} and \ref{snchi2}. It is clear from these plots that we can fit the data with a range of parameters. For the set of orbital parameters and masses listed in Table \ref{bestsn}, we measured precession rates from numerical integrations and calculated analytical precession rates (Table \ref{snrates}) for each satellite, where the analytical contributions consist of secular and $J_2$ calculations. The secular eigenfrequencies are $g_1$ $\sim$ 0.138 deg/day (or period of 7.2 years), $g_2$ $\sim$ 0.024 deg/day (period of 41.9 years), $f_1$ $\sim$ -0.161 deg/day (period of 6.1 years), and $f_2$ $\sim$ 0 deg/day. Overall, there is good agreement between analytical estimates and the rates observed from numerical integrations. Our analytical precession rates are affected by uncertainties in orbital parameters, mainly by those in semi-major axes. Although precession measurements can potentially provide powerful constraints on component masses and primary oblateness, the limited observational span prevents us from making stronger conclusions.

\def\arraystretch{1.4}
\begin{deluxetable}{l r r}[t]
\tablecaption{1994 CC: Best-Fit Parameters and Formal 1-$\sigma$ Errors\label{bestcc}} \\
	\startdata
		\hline \hline
		 & Beta (inner) & Gamma (outer) \\
		\hline
		Mass ($10^{10}$ kg) & 0.580 $\pm$ 0.331 & 0.091 $\pm$ 1.644 \\
		$a$ (km) & 1.729 $\pm$ 0.008& 6.130 $\pm$ 0.108  \\
		$e$ & 0.002 $\pm$ 0.015 & 0.192 $\pm$ 0.014 \\
		$i$ (deg) & 83.376 $\pm$ 11.158 & 71.709 $\pm$ 8.994 \\
		$\omega$ (deg) & 130.980 $\pm$ 43.647 & 96.229 $\pm$ 5.017 \\
		$\Omega$ (deg) & 59.209 $\pm$ 3.910 & 48.479 $\pm$ 4.741 \\
		$M$ (deg) & 233.699 $\pm$ 43.941 & 6.070 $\pm$ 6.187 \\
		$P$ (days) & 1.243 $\pm$ 0.0329 & 8.376 $\pm$ 0.404 \\
		\hline \hline
		 & \multicolumn{2}{r}{Alpha (central body)} \\
		\hline
		Mass ($10^{10}$ kg) & \multicolumn{2}{r}{25.935 $\pm$ 1.315}  \\
		$J_2$ & \multicolumn{2}{r}{0.014 $\pm$ 0.383} \\
		Pole Solution (deg) & \multicolumn{2}{r}{RA: -30.791 $\pm$ 3.910} \\
		 & \multicolumn{2}{r}{DEC: 6.624 $\pm$ 11.158}
		\enddata
\tablenotetext{}{The masses are listed in $10^{10}$ kg, $a$ is the semi-major axis, $e$ is the eccentricity, $i$ is the inclination, $\omega$ is the argument of pericenter, $\Omega$ is the longitude of the ascending node, $M$ is the mean anomaly at epoch, and $P$ is the period. These orbital elements are valid at MJD 54994 in the equatorial frame of J2000. Alpha's pole solution is given in right ascension (RA) and declination (DEC). This table lists formal 1-$\sigma$ statistical errors; see text for adopted 1-$\sigma$ uncertainties.}
\end{deluxetable}	

\def\arraystretch{1.4}
\begin{deluxetable*}{l r r r r r r r r r}
	  \tablecaption{1994 CC Precession Rates \label{ccrates}} \\
	  \startdata
	  	\hline \hline
	  	 & {\bf Beta (inner)} & & & & {\bf Gamma (outer)} & & & \\
	  	 & Analytical: & & & Numerical  & Analytical: & & & Numerical  \\
	  	 & Secular & $J_2$ & Total & Total  & Secular & $J_2$ & Total & Total  \\
	  	\hline
	  	$\dot{\omega}$ (deg/day)  & -0.612 & 0.392 & -0.219 & -0.058 & 0.135 & 0.005 & 0.139 & 0.131 \\
	  	$\dot{\Omega}$ (deg/day)  & -0.043 & -0.196 & -0.239 & -0.143 & -0.068 & -0.002 & -0.070 & -0.072 \\
	  	$\dot{\varpi}$ (deg/day)  & -0.655 & 0.196 & -0.459 & -0.202 & 0.067 & 0.002 & 0.069 & 0.061
	  \enddata
\tablenotetext{}{This table includes the argument of pericenter rate $\dot{\omega}$, the longitude of the ascending node rate $\dot{\Omega}$, and the longitude of pericenter rate $\dot{\varpi}$. For each satellite, the first 3 columns represent our analytical calculations and the fourth column shows our measured numerical rates.}
\end{deluxetable*}

\begin{figure*}[htbp]
		\centering
		\mbox{\subfigure{\includegraphics[scale=0.5]{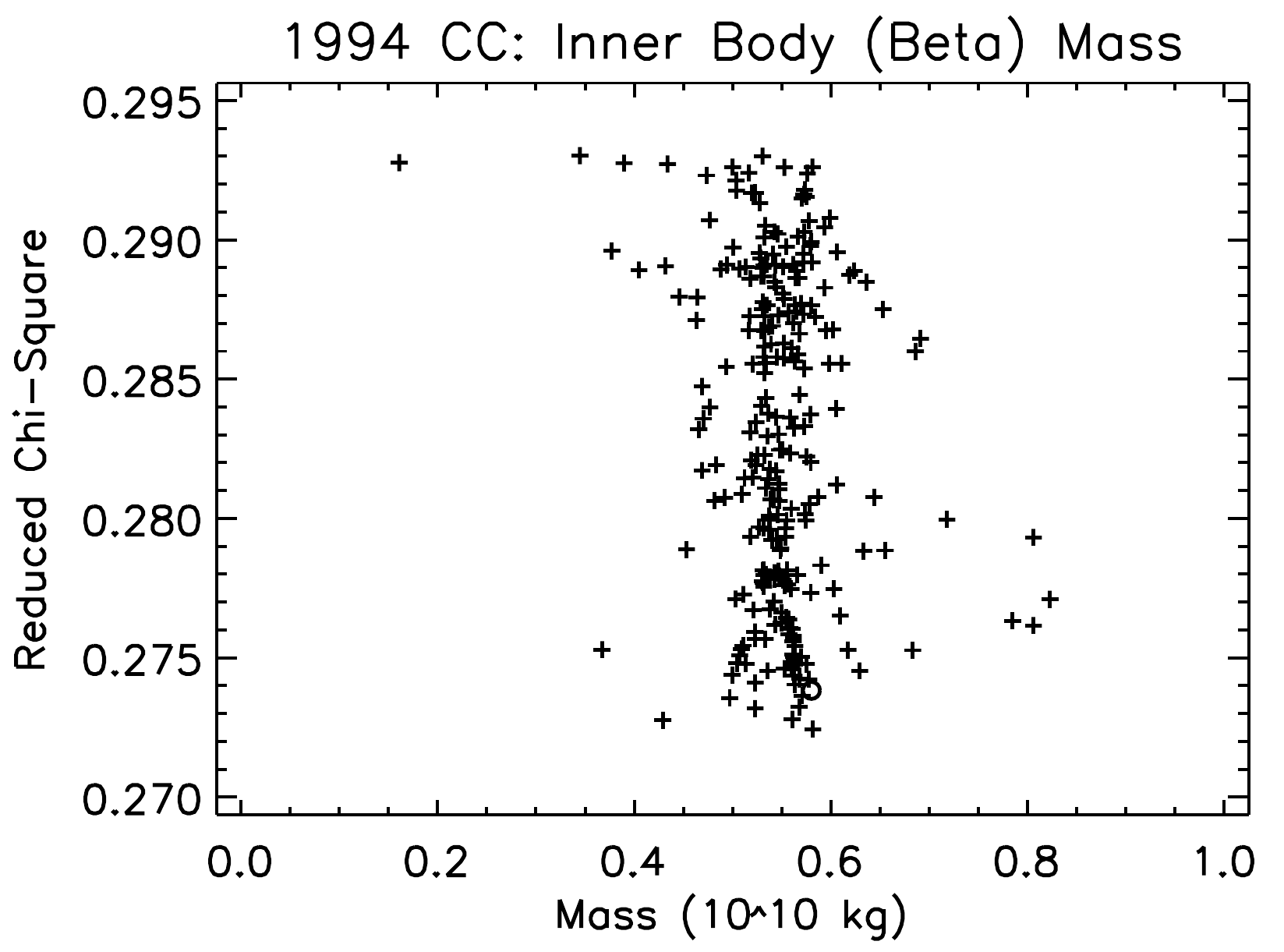}}\quad
		\subfigure{\includegraphics[scale=0.5]{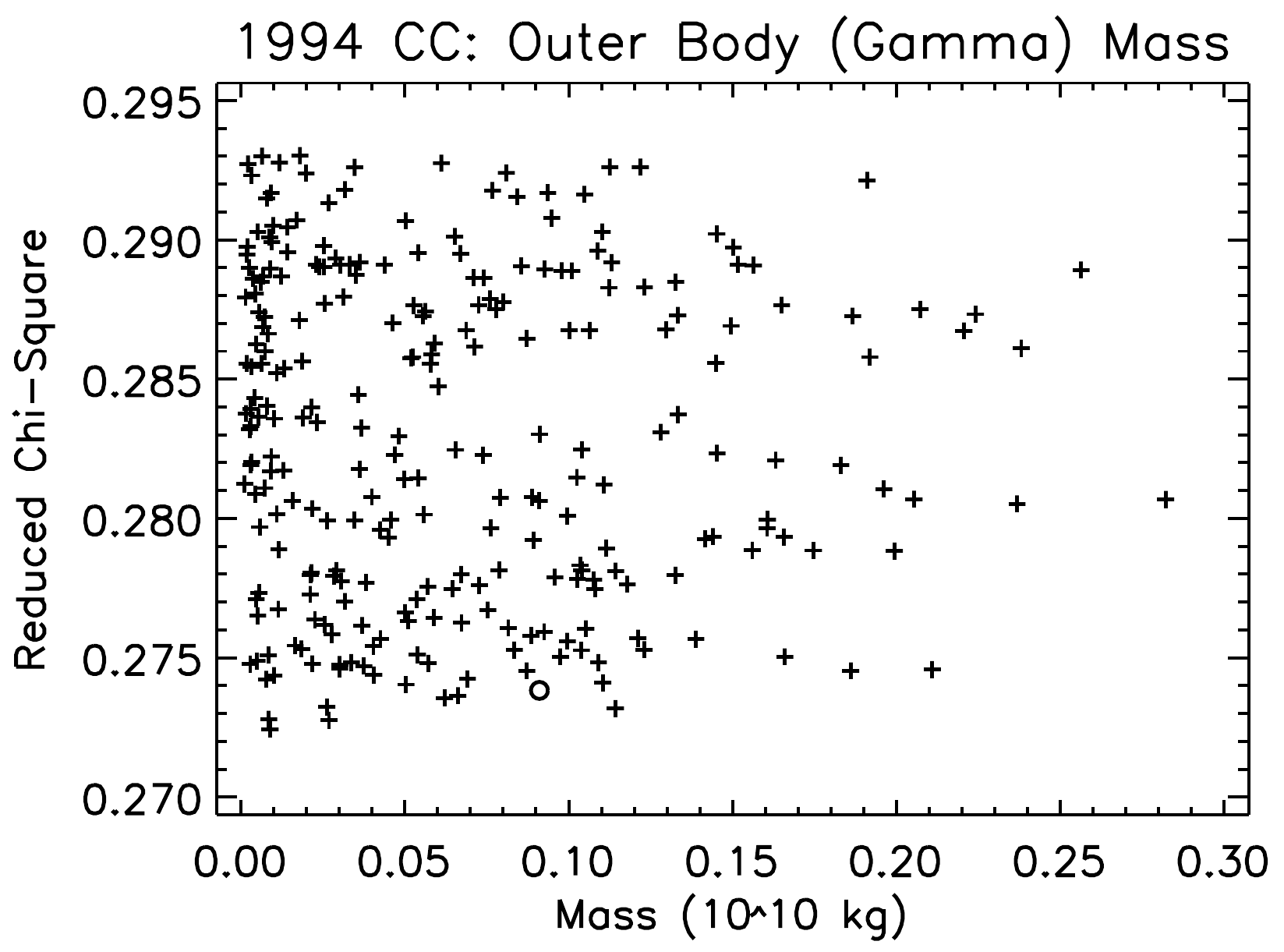}}}
		\mbox{\subfigure{\includegraphics[scale=0.5]{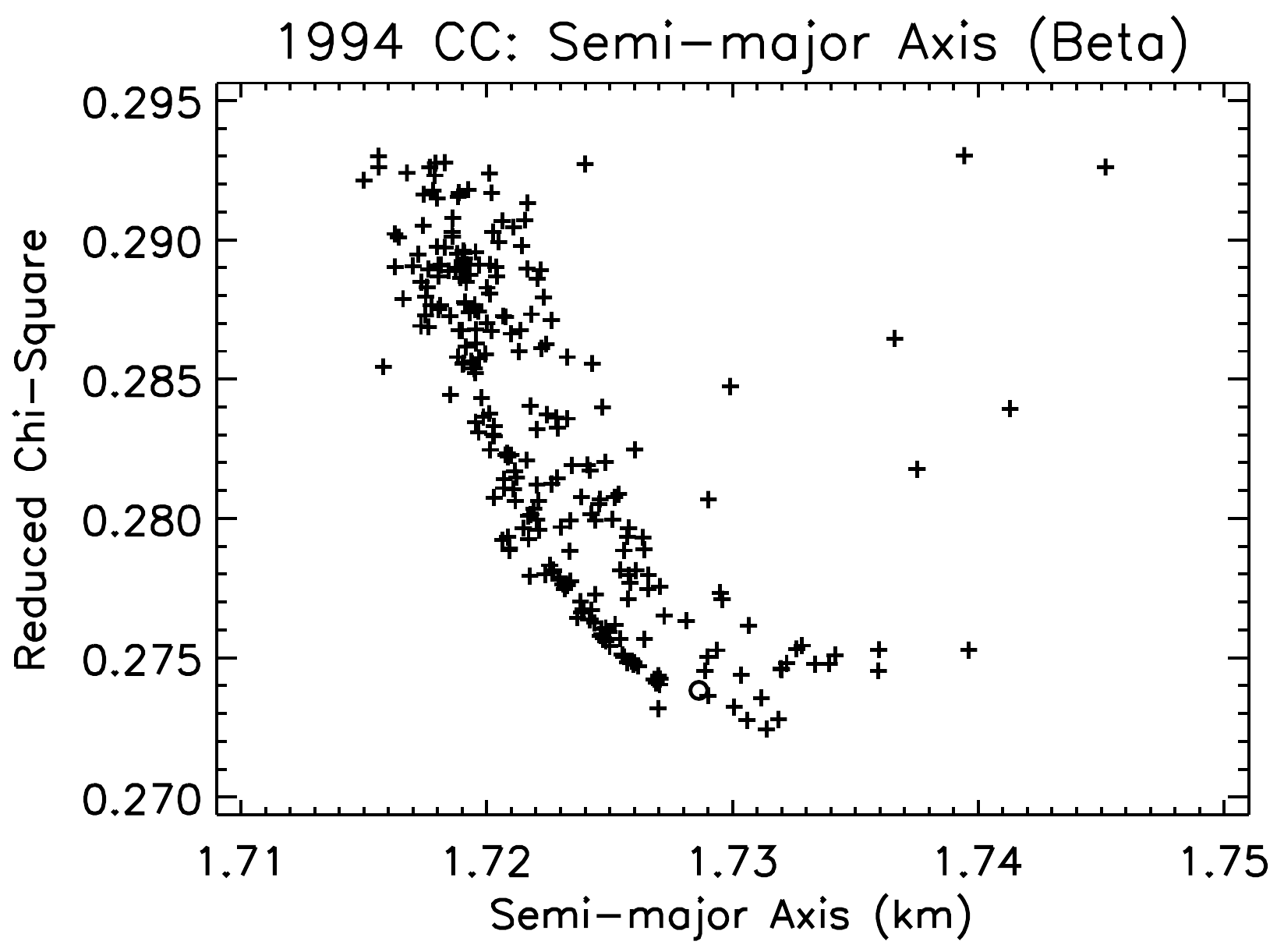}}\quad
		\subfigure{\includegraphics[scale=0.5]{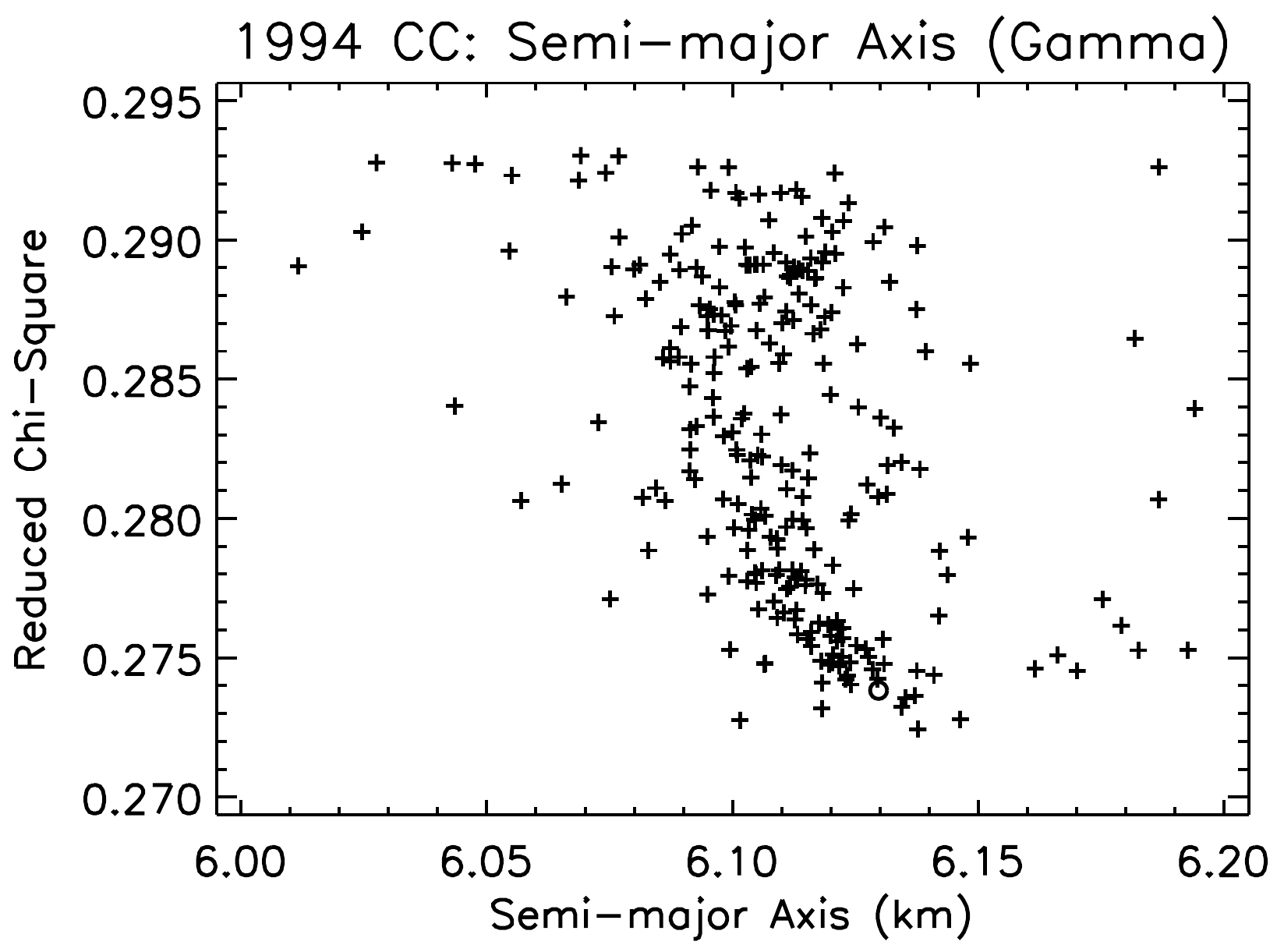}}}
		\mbox{\subfigure{\includegraphics[scale=0.5]{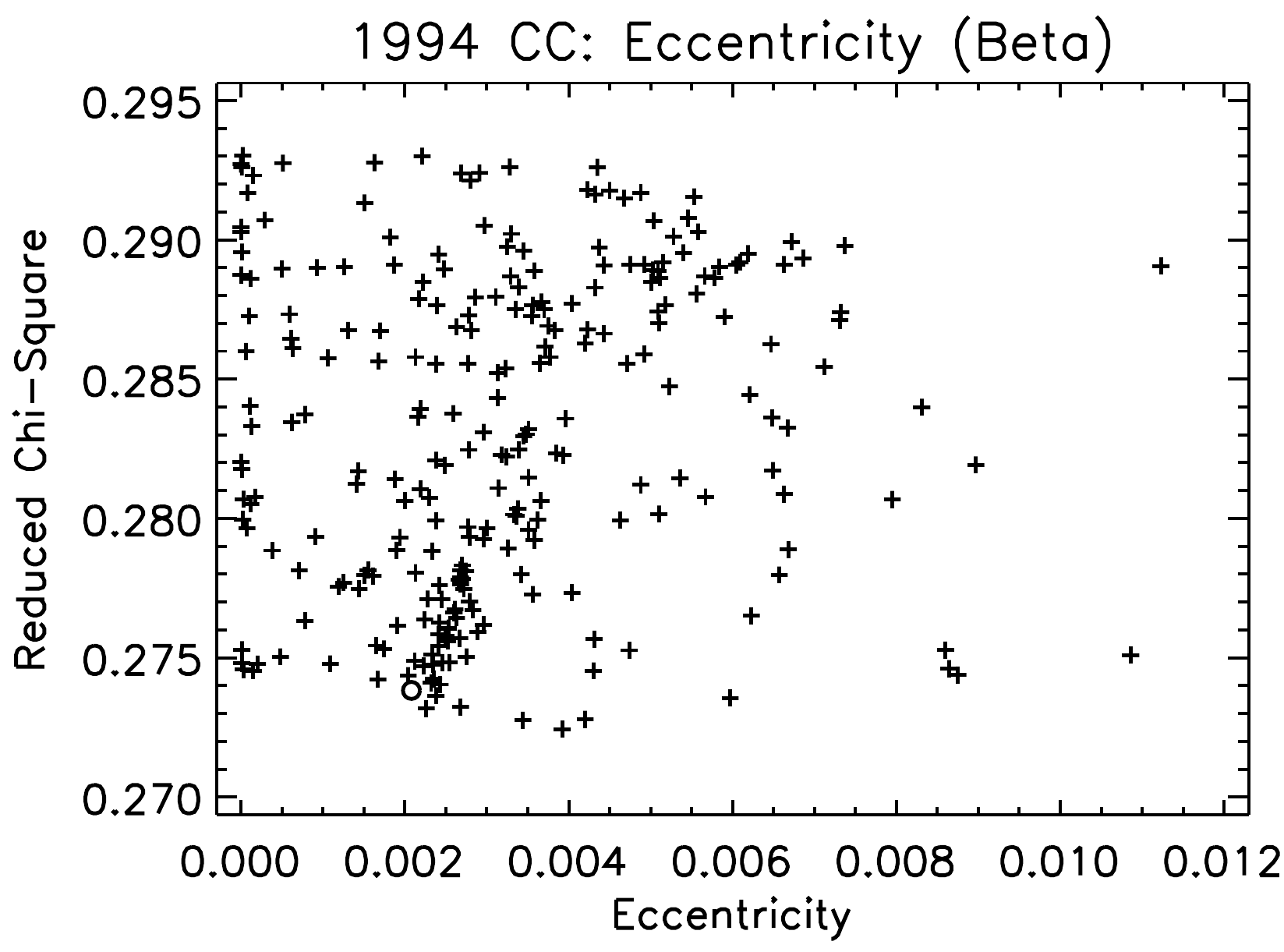}}\quad
		\subfigure{\includegraphics[scale=0.5]{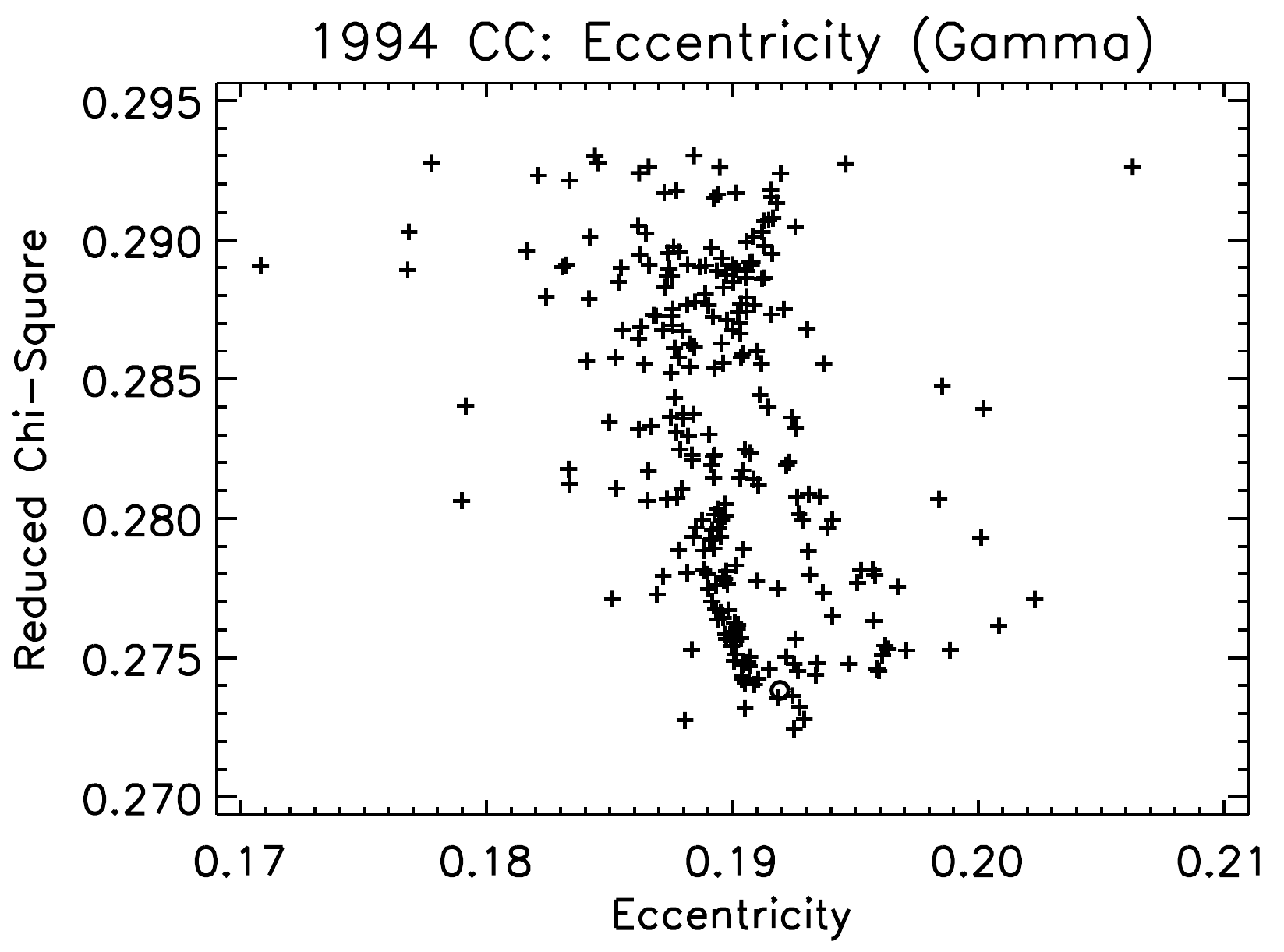}}}
		\caption{1994 CC: Reduced chi-square as a function of orbital parameters: mass, semi-major axis, and eccentricity. The points plotted here include the statistically equivalent best-fit orbital solutions that passed the constraints listed in Table \ref{constraints} and Section 2. Our adopted value is shown as an open circle, and solutions with lower chi-squares were ruled out on the basis of implausible $J_2$ and Beta/Gamma mass ratios.} \label{ccchi1}
\end{figure*}

\begin{figure*}[htbp]
		\centering
		\mbox{\subfigure{\includegraphics[scale=0.5]{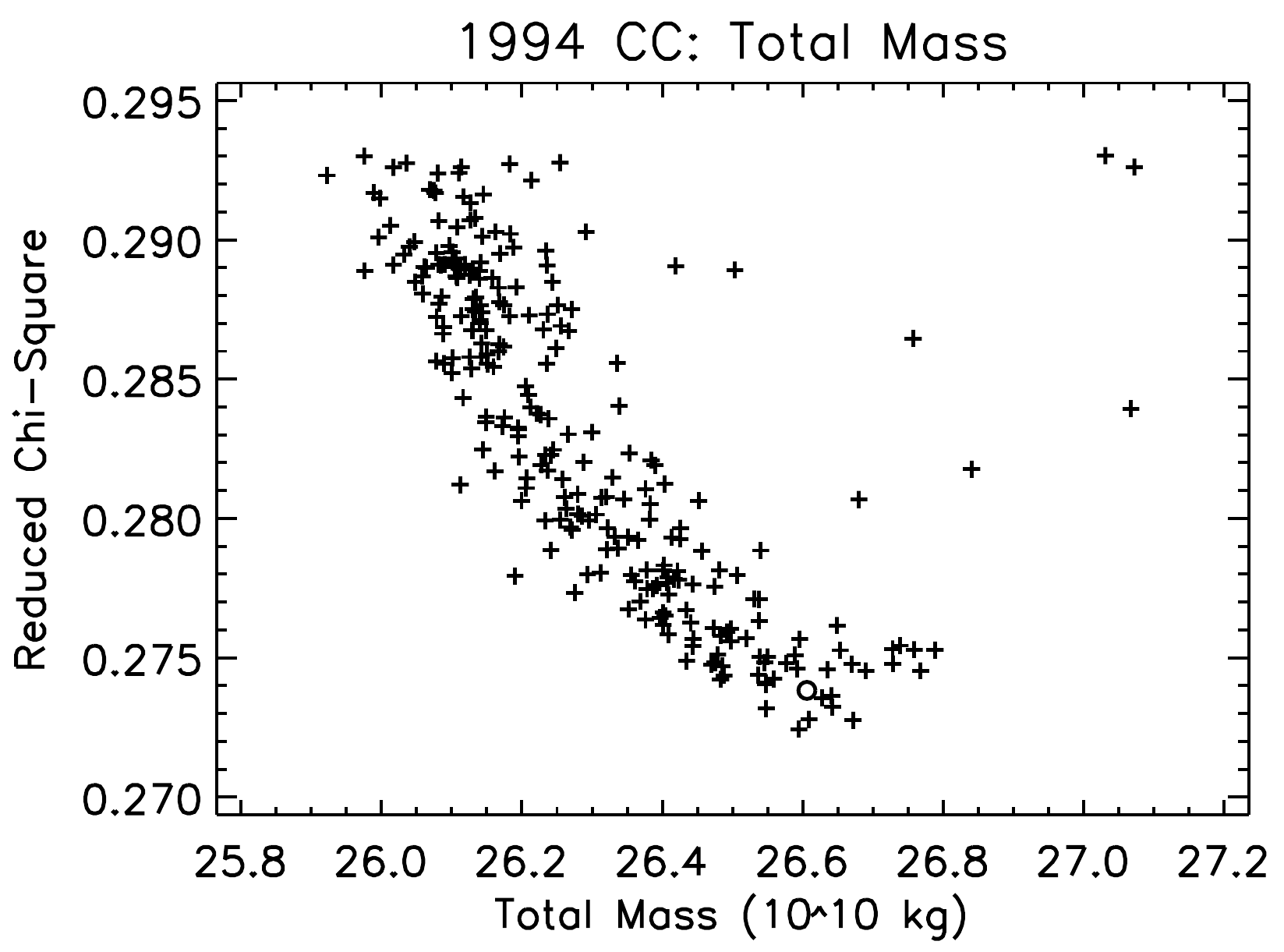}}\quad
		\subfigure{\includegraphics[scale=0.5]{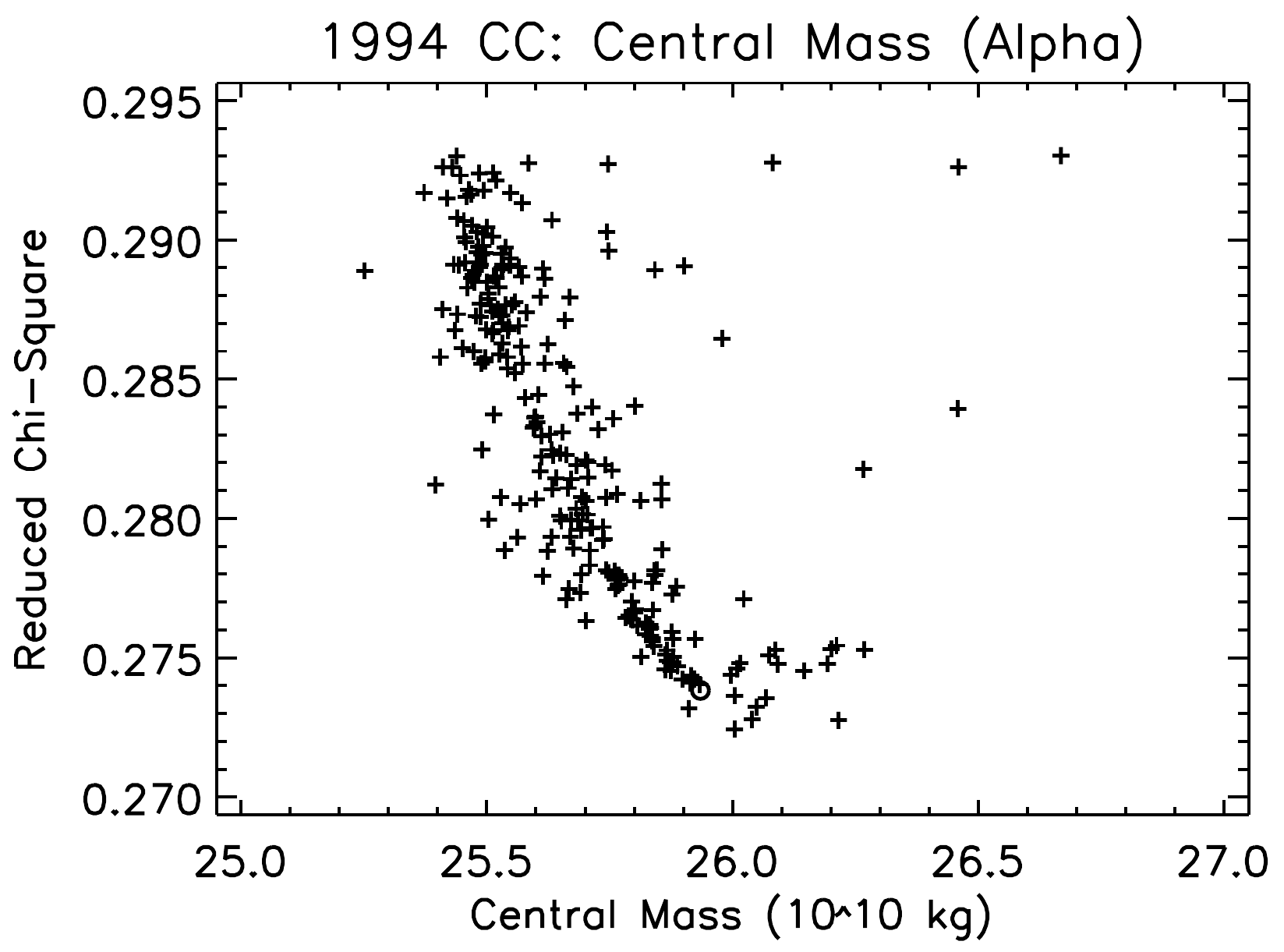}}}
		\mbox{\subfigure{\includegraphics[scale=0.5]{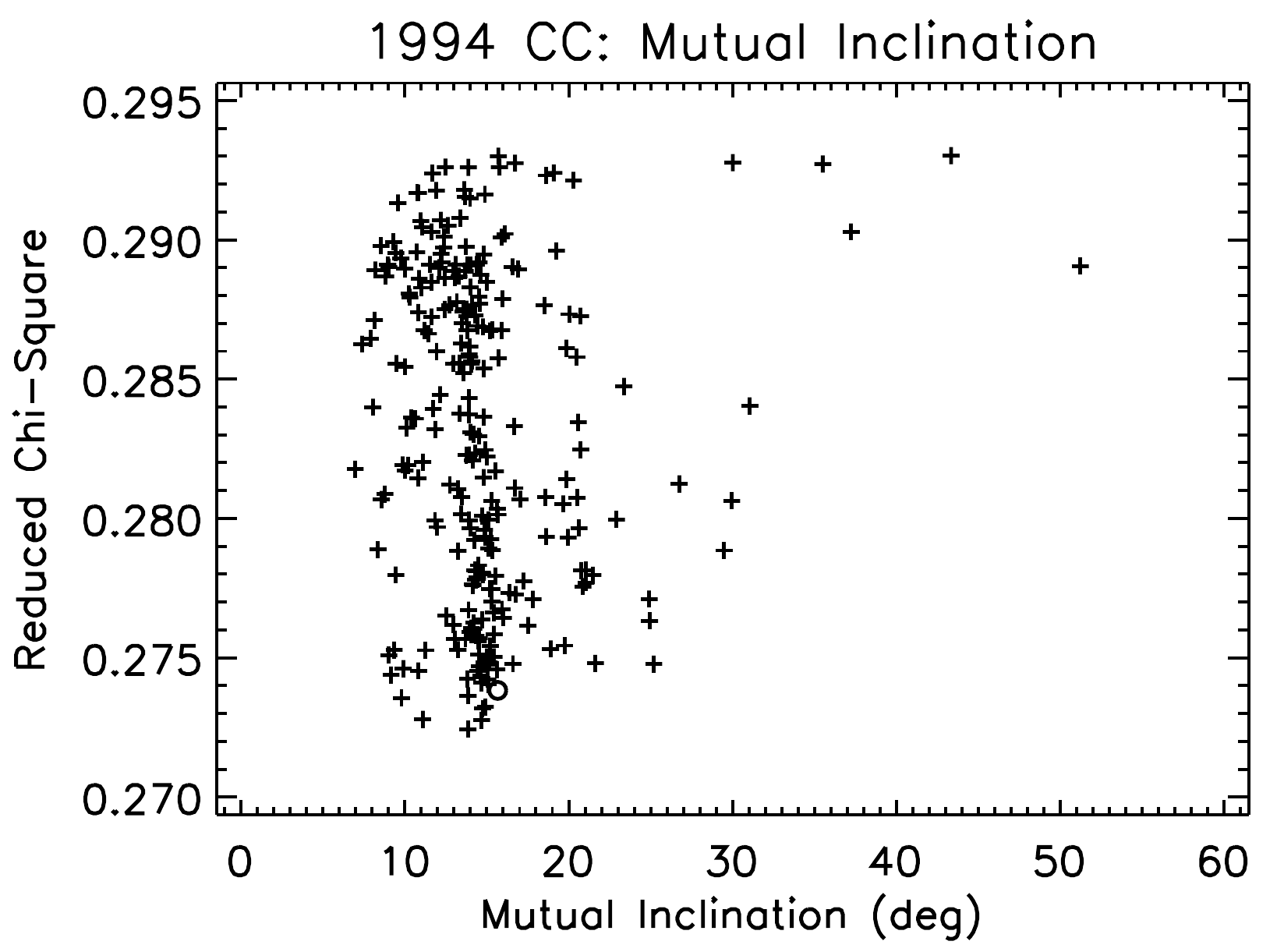}}\quad
		\subfigure{\includegraphics[scale=0.5]{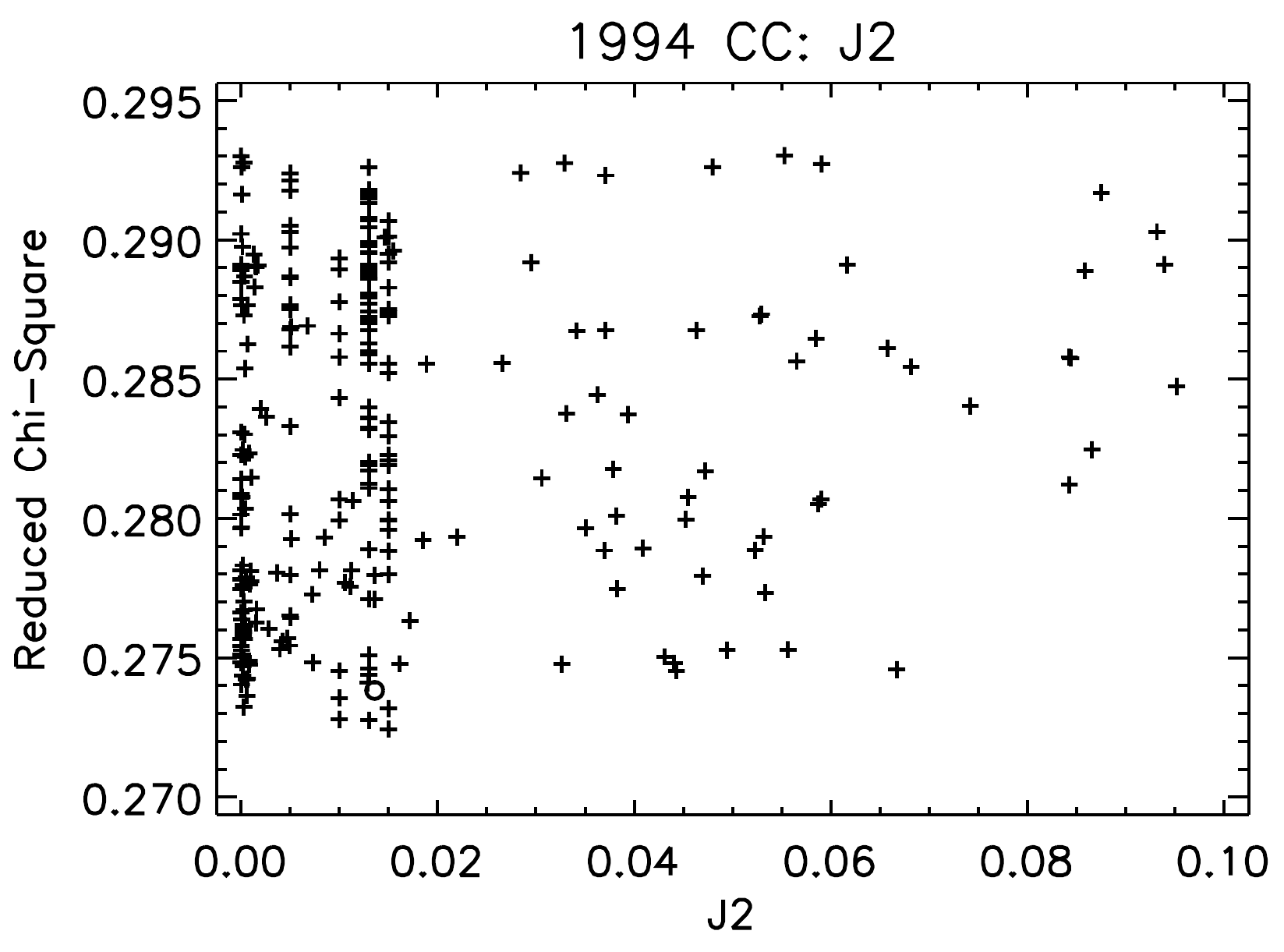}}}
		\caption{1994 CC: Reduced chi-square as a function of masses, mutual inclination, and $J_2$. The points plotted here include the statistically equivalent best-fit orbital solutions that passed the constraints listed in Table \ref{constraints} and Section 2. Our adopted value is shown as an open circle, and solutions with lower chi-squares were ruled out on the basis of implausible $J_2$ and Beta/Gamma mass ratios. There is structure seen in the $J_2$ plot due to fits where we constrained the $J_2$ parameter to be fixed at certain values: 0.005, 0.010, 0.013, and 0.015.} \label{ccchi2}
\end{figure*}

\subsection{1994 CC} 

	The best-fit parameters at MJD 54994.0 yield a reduced chi-square of 0.27 (Table \ref{bestcc}) with DOF = 364. While there are orbital fits with lower chi-squares, the particular solution described in this section and Table \ref{bestcc} has the most plausible combination of Beta/Gamma mass ratio and $J_2$ value. See Section 3.3 for discussion regarding masses and $J_2$. As is the case for 2001 SN263, some of the formal 1-$\sigma$ errors in Table \ref{bestcc} are likely to be underestimates of the actual uncertainties in the parameter values, and here we list our adopted 1-$\sigma$ uncertainties from examining our best-fit solutions along with parameter values with guard digits. The mass of Alpha is $25.935 \pm 1 \times 10^{10}$ kg, Beta is $0.580_{-0.5}^{+0.3} \times 10^{10}$ kg, and Gamma is $0.0911_{-0.09}^{+0.20} \times 10^{10}$ kg. We estimate a density of $\sim$1.98 g/cm$^3$ for Alpha, given its preliminary size estimate from radar images. If we apply Alpha's density to the range of Beta and Gamma masses that are acceptable (within uncertainties), we find that their equivalent radii can range from 46 - 102 m for Beta and 11 - 71 m for Gamma. Beta's orbit is nearly circular at $0.002_{-0.002}^{+0.009}$ with a semi-major axis of $1.729 \pm 0.02$ km and Gamma's orbit has an eccentricity of $0.192_{-0.022}^{+0.015}$ and a semi-major axis of $6.13_{-0.12}^{+0.07}$ km. The orbital periods of Beta and Gamma are $1.243 \pm 0.1$ and $8.376 \pm 0.5$ days, respectively. This fit converged to a $J_2$ value of $0.014_{-0.014}^{+0.050}$, where Alpha's pole was assumed to be aligned with Beta's orbit normal. From this orbital solution, we find a significant mutual inclination of $\sim$16 degrees. We estimate angular uncertainties of $\sim$20 degrees for Beta's orbit pole and $\sim$10 degrees for Gamma's orbit pole.

	We show our statistically equivalent (within 1-$\sigma$ level) best fits in Figures \ref{ccchi1} and \ref{ccchi2}. For 1994 CC, we also compared precession rates (Table \ref{ccrates}) between our numerical integrations and analytical calculations and find good agreement for the outer body, Gamma. There is appreciable disagreement for Beta, which will be discussed in Section 3.4. The secular eigenfrequencies are $g_1$ $\sim$ 0.016 deg/day (or period of 60.1 years), $g_2$ $\sim$ 0.070 deg/day (period of 14.1 years), $f_1$ $\sim$ 0 deg/day, and $f_2$ $\sim$ -0.086 deg/day (period of 11.4 years). As is the case for 2001 SN263, our estimates of precession rates are affected by uncertainties in orbital parameters.

\subsection{Masses and $J_2$}

	In the case of 2001 SN263, we are limited in our ability to measure the masses of the satellites because the majority of our low chi-square fits with reasonable mass ratios ($m_{\rm Beta} > m_{\rm Gamma}$) have a lower $J_2$ coefficient ($\sim$0) than the expected value ($\sim$0.016) from shape model estimates (Nolan et al., in prep.). Our fits with reasonable mass ratios typically occurred in integrations where we allowed $J_2$ to float as a parameter in our least-squares minimization routine, which usually resulted in a very low and unlikely $J_2$ value of 0. When we attempted to fix $J_2$ at a range of values, including our best estimate of $J_2$ from shape model estimates, our low chi-square solutions typically had implausible mass ratios where Gamma was more massive than Beta. Since our best fit shown in Table \ref{bestsn} has a reasonable value of $J_2$, it is possible that our mass for Beta is an underestimate of its true value. For 1994 CC, we also found orbital solutions where $J_2$ was driven to 0, but there was a greater fraction of fits with both reasonable mass ratios and $J_2$ values.

	Nevertheless, in both systems we find a wide range of $J_2$ values (Figures \ref{snchi2} and \ref{ccchi2}) that produce orbital solutions with low chi-square results. As a result, we cannot constrain $J_2$ well with the current set of observations. One possibility is that there is a degeneracy between satellite mass (in particular, the outer satellite) and $J_2$, as suspected for Haumea by Ragozzine \& Brown (2009). This behavior can be explained by examining our precession rates for the inner body. For 2001 SN263, we expect that both the mass of the outer satellite and the $J_2$ of the primary produce an advance of the inner body's longitude of pericenter (Table \ref{snrates}) and for 1994 CC, the effects of the mass of the outer satellite and the $J_2$ of the primary are opposite (Table \ref{ccrates}), which can happen at certain phases of the secular evolution. It is difficult to apportion the respective contributions of the two quantities, $J_2$ and the mass of the outer body, to the precessional effects of the inner body. Further support for this idea comes from a comparison of the outer body's mass and $J_2$ values for the ensemble of our valid solutions. We find that the two quantities are appreciably anti-correlated (r = -0.48) for 2001 SN263, with the sum of the two contributions matching the measured rates. In the case of 1994 CC, a mild positive correlation was found between the outer body's mass and $J_2$ (r = 0.19).

Studies on near-Earth binary 1999 KW4 by Ostro et al. (2006) and Scheeres et al. (2006) have demonstrated some of the complex dynamics that can result from interactions between asymmetrically-shaped components. In particular, Cuk \& Nesvorny (2010) found that the libration of an elongated asteroid satellite can result in a negative apsidal precession rate.  Since this type of libration-induced precession may take place but is not included in our model, we speculate that the minimization procedure artificially lowers the value of the parameter $J_2$, which would normally result in a positive apsidal precession rate. Including libration-induced precession requires knowledge of the component shapes that is not available at this time; it is therefore beyond the scope of this paper.

\
\subsection{Precession Rates: Numerical and Analytical Comparisons}

There is good agreement in the precession rates for 2001 SN263 (Table \ref{snrates}) between those measured from numerical integrations and our analytical estimates, described in Section 2.3. For 1994 CC, we see good agreement between our analytical and numerical precession rates for the outer body, Gamma (Table \ref{ccrates}). For the inner body (Beta), whose orbit lies in Alpha's equatorial plane, there is appreciable disagreement between the total values for the precession rates. We attribute this disagreement to the non-secular terms in the disturbing function; due to the infinite number of short-period terms in the disturbing function, there are differences between the results of our numerical integration and the predictions of secular analytical theory. We verified this by calculating the precession rates from numerical integrations with an outer body mass of near zero, which meant that any precession of the inner body must be due to $J_2$ only. Indeed, our inner body's precession rate ($\dot\varpi \approx$ 0.197 deg/day) from numerical integrations with essentially no outer body closely matched the expected analytical precession ($\dot\varpi \approx$ 0.196 deg/day) due to $J_2$ only. Thus, it is likely that any disagreement between our numerical and analytical values for Beta can be credited to non-secular terms in the disturbing function. We suspect that the reason we see this discrepancy between numerical and analytical rates for 1994 CC's Beta and not for the other bodies in these triple systems is the fact that the perturber in this case is Gamma, whose orbital plane is both eccentric as well as inclined to Alpha's equatorial plane and Beta's orbital plane. 

\subsection{Mean-Motion Resonance}

\begin{figure}[t!]
	\centering
	\includegraphics[scale=0.35]{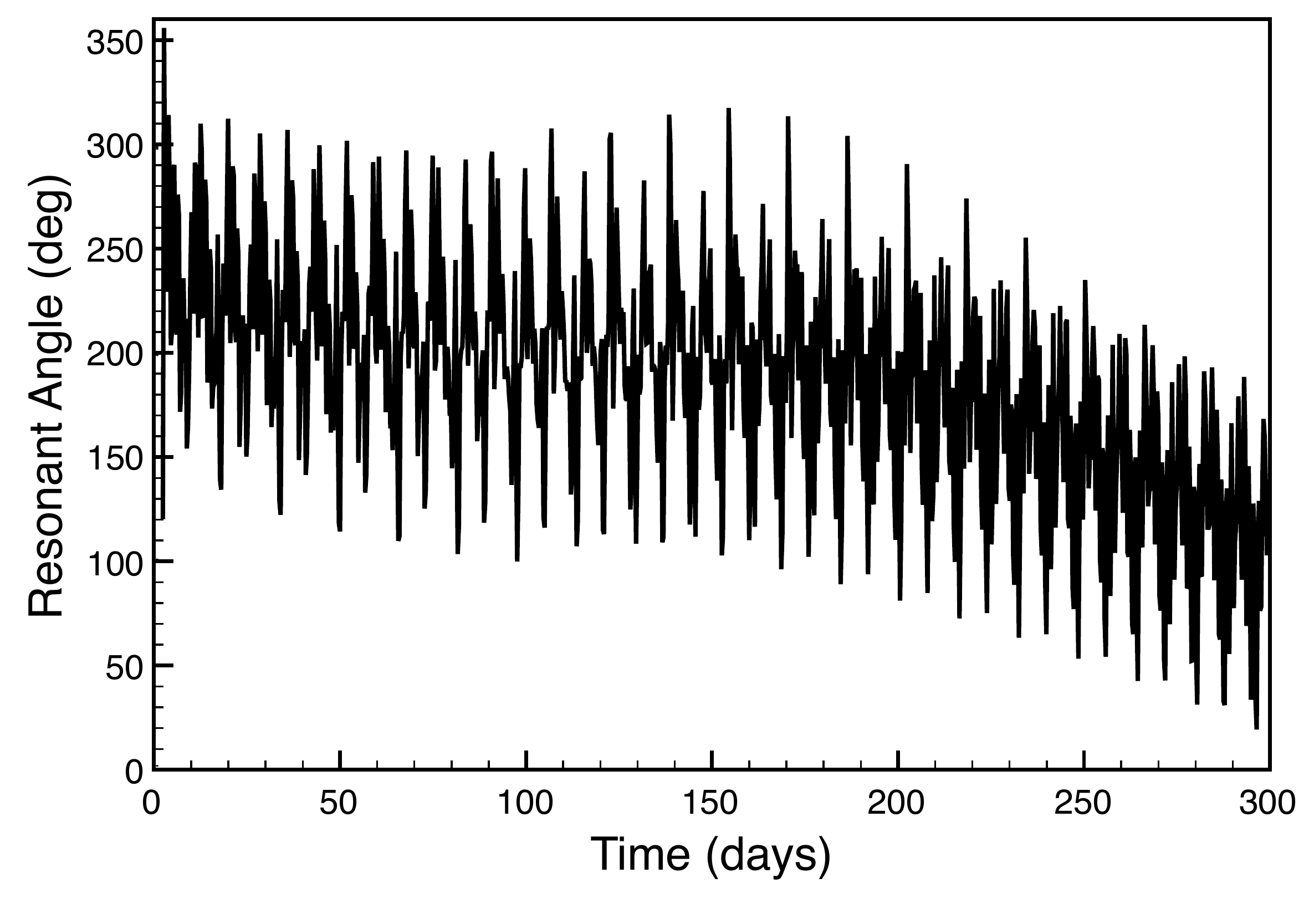}
	\caption{This figure shows a sample resonant argument for 1994 CC, where there initially appears to be libration (oscillating motion) followed by circulation through all angles.} \label{cclibrat}
\end{figure}

	The satellites of 2001 SN263 have an orbital period ratio near 9:1 using their nominal orbital elements. However, we did not find any librating resonant arguments over long, secular timescales for either system. We found cases where a resonant argument appeared to librate initially and then circulated thereafter; an example of this is shown in Figure \ref{cclibrat} for 1994 CC. In this example, we see a distinct onset of circulation as a result of the secular precession of the longitudes of pericenter and the ascending node. Both of these longitudes are changing over time (see Table \ref{ccrates}), causing the resonant argument in this situation to cease its apparent libration and to start circulating over all possible angles.
	
\section{Orbital Origin and Evolution}

It is interesting to find a non-zero eccentricity ($\sim$0.19) for 1994 CC's outer body, Gamma. Furthermore, our best orbital solutions yielded significant mutual inclinations (2001 SN263: $\sim$14 degrees, 1994 CC: $\sim$16 degrees) between the orbital planes of the satellites. Other known systems with a significant inclination include the main-belt asteroid triple (45) Eugenia, whose satellites have been reported to be inclined 9 and 18 degrees with respect to the primary's equator (Marchis et al. 2010) and the dwarf planet Haumea, whose two satellies have a mutual inclination of 13.41 degrees (Ragozzine \& Brown 2009). 

Possible mechanisms to excite orbital eccentricity and inclination include Kozai resonance, evection resonance, mean-motion resonance crossings, close planetary encounters, and a combination of these effects. In the subsequent sections, we calculate the eccentricity and inclination damping timescales and investigate each of these possible mechanisms.

\subsection{Eccentricity and Inclination Damping Timescales}

Tidal damping of a synchronous satellite's orbit is a competing process between tides raised on the satellite by the primary (causing the eccentricity to decay) and tides on the primary due to the satellite (causing both the semi-major axis and eccentricity to grow). Tidal dissipation in the satellite's interior due to the primary can heat the satellite and circularize its orbit. The corresponding eccentricity damping timescale $\tau_{\rm e,damp}$ (Murray \& Dermott 1999) is:
\begin{equation}
	\tau_{\rm e,damp} = \dfrac{4}{63}\dfrac{m}{M}\left(\dfrac{a}{r}\right)^{5}\left(\dfrac{\tilde{\mu}_sQ_s}{n}\right)
\end{equation}
and the competing timescale for tides raised on the primary is:
\begin{equation}
	\tau_{\rm e,excite} = \dfrac{16}{171}\dfrac{M}{m} \left(\dfrac{a}{R}\right)^5 \left(\dfrac{\tilde{\mu}_c Q_c}{n}\right)
\end{equation}	
where $M$ is the central body's mass, $m$ is the mass of the satellite, $a$ is the semi-major axis, $R$ is the central body's mean radius, $r$ is the satellite's mean radius, $\tilde{\mu}$ is a measure of a body's rigidity, $Q$ is the tidal dissipation function, and $n$ is the mean motion. The subscripts $c$ and $s$ represent the central body and the satellite, respectively. The effective rigidity $\tilde{\mu}$, a unitless quantity as defined in Murray \& Dermott (1999), is uncertain but can be estimated for gravitational aggregates (``rubble piles") as shown in Goldreich \& Sari (2009): we find $\tilde{\mu}_s$ $\sim$ 3.1 $\times 10^{6}$ for Gamma ($r \sim$ 45 m). The factor $Q$ is even harder to determine, but available evidence suggests that Q $\sim$ $10^2$ for monoliths and potentially smaller for rubble piles. With these values as well as our best-fit solution (Table \ref{bestcc}) for 1994 CC, the eccentricity damping timescale for Gamma, if synchronously rotating, is $\tau_{\rm e,damp} \sim 10^{13}$ years. There is evidence that Gamma is, in fact, not synchronously rotating (Brozovic et al. 2010); however, even if Gamma is synchronous, we have shown that its eccentricity cannot damp on billion-year timescales because it is too small and distant.

Assuming identical composition and $Q$ values for Alpha and Gamma, the ratio of the timescales of eccentricity damping and excitation for rubble piles are
\begin{equation}
	\dfrac{\tau_{\rm e,damp}}{\tau_{\rm e,excite}} = \dfrac{19}{28}
\end{equation}
which shows that tides will always damp the eccentricity for a synchronously rotating satellite regardless of the system's size and mass ratios (Goldreich \& Sari 2009). The tidal despinning timescale, which is the time for a satellite to reach synchronous rotation with its central body, can be estimated for the satellites in both systems. For 2001 SN263, Gamma (inner) has a despinning timescale on the order of $10^5$ years and Beta's (outer) timescale is on the order of $10^9$ years. For 1994 CC, Beta (inner) has a timescale on the order of $10^7$ years and Gamma's (outer) timescale is on the order of $10^{11}$ years. 2001 SN263 Gamma and 1994 CC Beta are therefore plausible synchronous rotators. If 1994 CC Beta was once closer to Alpha, its despinning timescale would be accordingly shorter.

The inclination damping timescale $\tau_{\rm I,damp}$ is related to the eccentricity damping timescale (Yoder \& Peale 1981):
\begin{equation}
	\tau_{\rm I,damp} = \frac{7}{4}\tau_{\rm e,damp}\left(\dfrac{\sin I}{\sin\epsilon}\right)^2
\end{equation}
where $I$ is the inclination of the orbit with respect to the central body's equator and $\epsilon$ is the obliquity of the satellite's spin axis relative to the orbit normal. To arrive at an estimate of this timescale, we assume $\epsilon \sim$ 1 deg and find that $\tau_{\rm I,damp}$ is on the order of $10^{15}$ years for 1994 CC Gamma. These long eccentricity and inclination damping timescales indicate that tides cannot damp out $e$ and $I$ on timescales comparable to possible excitations of the system, which we now examine, nor the dynamical and collisional lifetimes of the system.

\subsection{Kozai Resonance}

Kozai resonance (Kozai 1962; Murray \& Dermott 1999) is an angular momentum exchange process between a satellite's eccentricity and inclination that takes place under the influence of a massive outer perturber, provided that the relative inclination between the orbits of the satellite and the perturber exceeds a limiting value.

When considering the outer satellite as the perturber, we find that the Kozai process is not active, as the mutual inclinations do not exceed the required value. For instance, the limiting Kozai inclination corresponding to 2001 SN263's semi-major axis ratio ($a_{\rm inner}/a_{\rm outer} \sim$ 0.23) can be estimated as $\sim$37.5 degrees (Kozai 1962). The currently observed mutual inclination between the orbital planes of 2001 SN263 Beta and Gamma is only $\sim$14 degrees.

We also consider the case with the Sun as the massive, outer perturber (as in Perets \& Naoz 2009). The respective inclinations between the Sun's apparent orbit around Alpha and the orbits of the satellites are $\sim$8 and $\sim$17 degrees for 2001 SN263 Beta and Gamma and $\sim$76 and $\sim$61 degrees for 1994 CC Beta and Gamma. Consequently, while 2001 SN263 does not satisfy the limiting Kozai inclination (in this case estimated as $\sim$39.2 degrees), 1994 CC can be prone to Kozai oscillations. From numerical integrations corresponding to our best-fit parameters for 1994 CC, we do not see either satellite in Kozai resonance with the Sun. In integrations where we have removed one satellite, such that there were only three bodies total (Alpha, Beta or Gamma, and the Sun), we saw that Gamma is particularly susceptible to Kozai resonance.  We observed the telltale signs: strong coupled oscillations of eccentricity and inclination with the relevant period of $\sim$56 years, conservation of the Delaunay quantity $H_k=\sqrt{(1-e^2)}\cos I$, and libration of the argument of pericenter. The differences between the two cases is that when the inner body is present, it causes the outer body's argument of pericenter to precess too fast for libration to occur. When both satellites are present, Kozai oscillations due to the Sun are suppressed. Therefore, Kozai interactions are not likely to explain the observed mutual inclinations and eccentricity.

\subsection{Evection Resonance}

Evection resonance can occur when the satellite's orbit around Alpha has a longitude of pericenter $\varpi$ that precesses at the same rate as the Sun's apparent mean longitude $\lambda_s$ with respect to Alpha. This results in libration of the argument $\phi = 2\lambda_s - 2\varpi$, and can produce a change in the eccentricity of the orbit (Touma \& Wisdom 1998). 

None of the satellites are currently in a configuration where the evection resonance can affect their evolution. However, as the semi-major axes change over time due to tidal or radiation effects, the precession rates evolve and Beta or Gamma can cross the evection resonance. However, as in the case of Kozai resonance, mutual perturbations will likely suppress the evection resonance or prevent it from being active for very long. Therefore, evection interactions are not likely to explain the observed mutual inclinations and eccentricity.

\subsection{Passage Through Mean-Motion Resonances}

Tidal evolution and binary YORP (BYORP; Cuk \& Nesvorny 2010), which affects synchronous satellites, can change the orbits and lead to mean-motion resonances in the system. The outer bodies in both triple systems cannot be directly affected by BYORP as there is evidence that they are not in synchronous rotation (Nolan et al. 2008b; Brozovic et al. 2010). For the inner bodies where the synchronization condition is more likely to be met, it is possible that BYORP will cause migration of their orbits. This can result in an increase or decrease of the semi-major axis, depending on the satellite's shape. Previous work have shown that outward migration can cause an increase in the orbit's eccentricity (Cuk \& Burns 2005), but inward migration can lead to a decrease in free eccentricity and is thought to be the most likely result of its evolution (Cuk \& Nesvorny 2010). On the contrary, other studies (McMahon \& Scheeres 2010a; McMahon \& Scheeres 2010b) find that over long timescales, the eccentricity changes in the opposite direction of the semi-major axis. As for tidal evolution, this process will change the semi-major axis as $da/dt \propto a^{(-11/2)}$ (Murray \& Dermott 1999), leading to converging orbits. 

Whether due to tidal or BYORP evolution, a change in semi-major axes can lead to capture or passage through mean-motion resonances, which in turn can excite eccentricity and inclination (Dermott et al. 1988; Ward \& Canup 2006). Such processes may therefore be responsible for the observed mutual inclinations and eccentricity.

\subsection{Close Planetary Encounters}

We tested the possibility that 1994 CC Gamma's nonzero eccentricity and both systems' mutual inclinations are due to close encounters with terrestrial planets. We performed systematic numerical simulations of the triple asteroid system and an Earth-sized body with over 5,000 permutations of orbital elements: inclinations with respect to Earth's equator, longitudes of the ascending node, and mean anomalies for both satellites. We started our simulations with equatorial, coplanar and circular orbits for Beta and Gamma, and integrated various close encounter distances up to 60 $R_{\Earth}$ with an encounter velocity $v_\infty$ of 12 km/s (typical for Earth-crossing asteroids). We validated this procedure by comparing with results from Bottke \& Melosh (1996) for a binary system.

\begin{figure}[bhtp]
	\centering
	\includegraphics[scale=0.35]{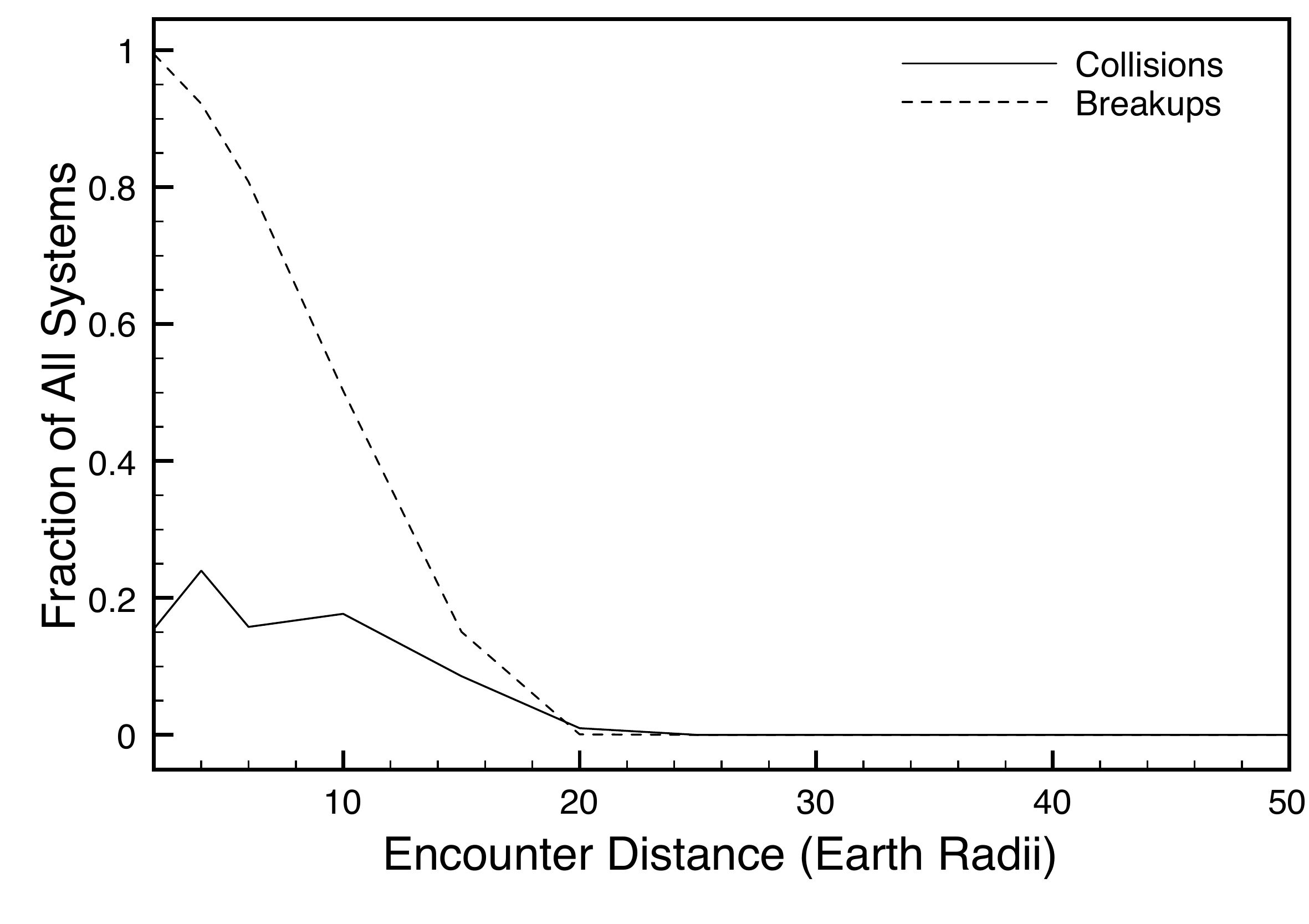}
	\includegraphics[scale=0.35]{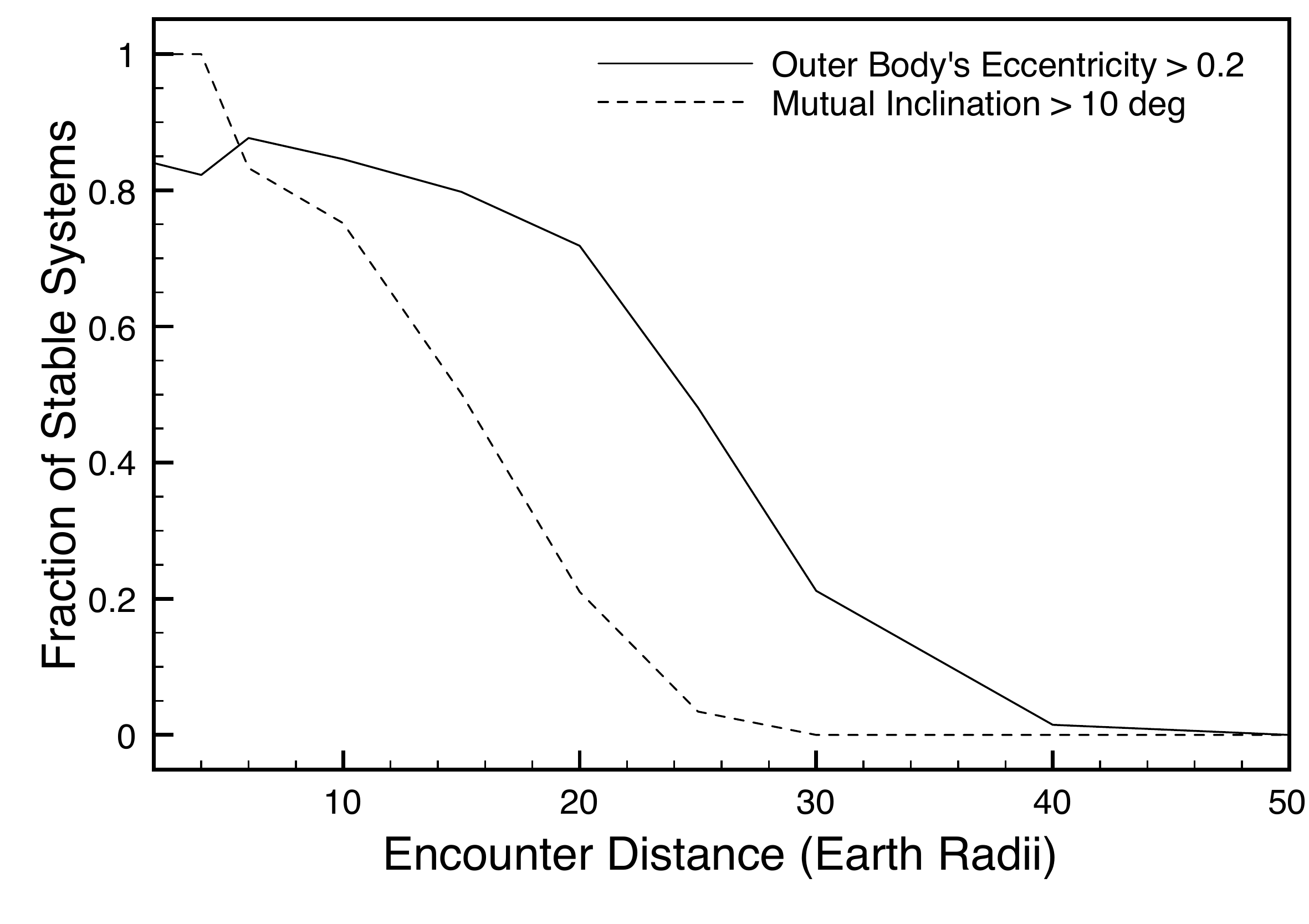}
\caption{2001 SN263: These figures show the close encounter statistics for the onset of instability (top) and excitation (bottom). The top diagram shows the fraction out of all systems; the bottom diagram shows the fraction out of only stable systems, defined as those with no collisions and ejections. \label{snflyby1}}
\end{figure}

\begin{figure}[bhtp]
	\centering
    \includegraphics[scale=0.35]{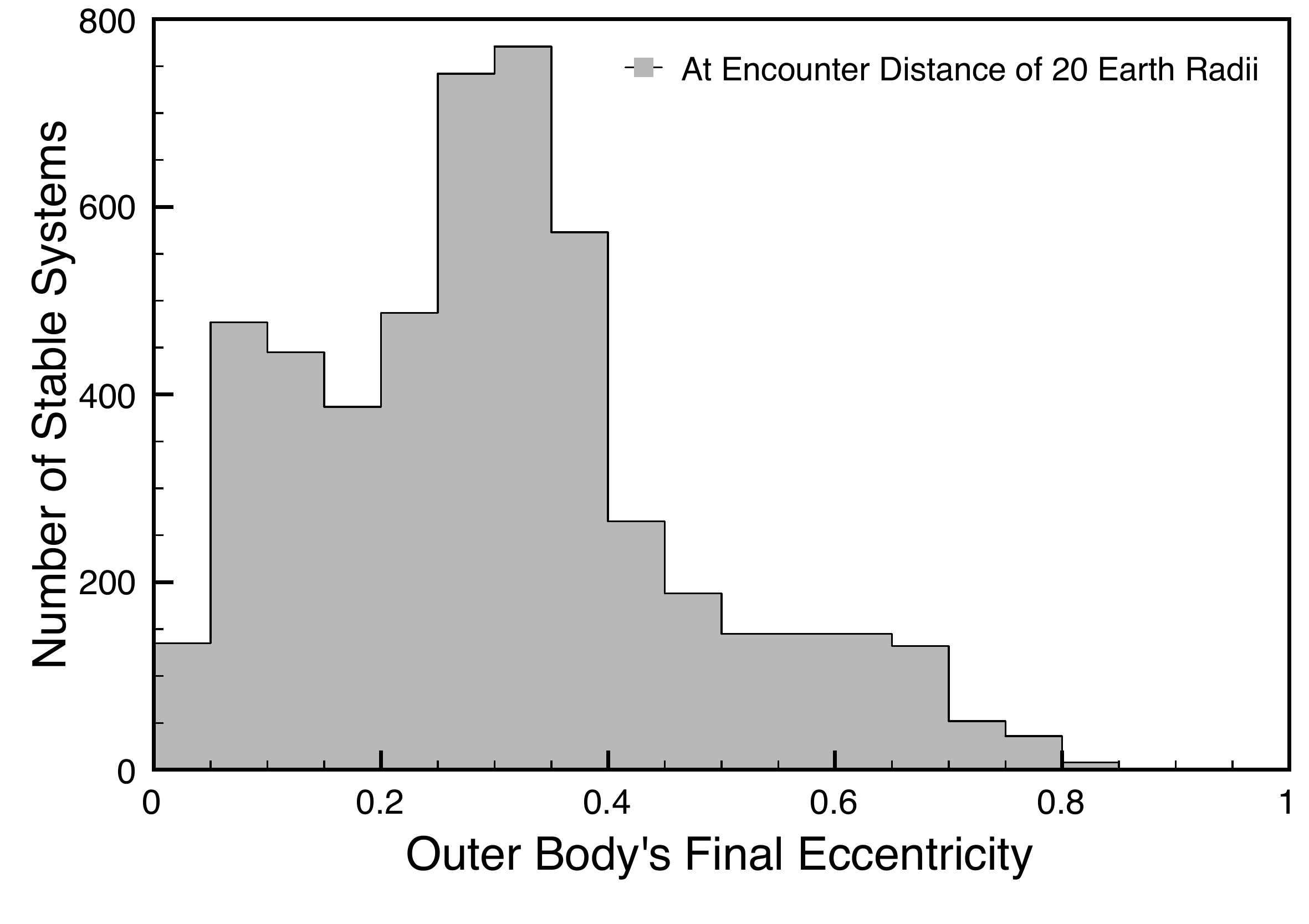}
    \includegraphics[scale=0.35]{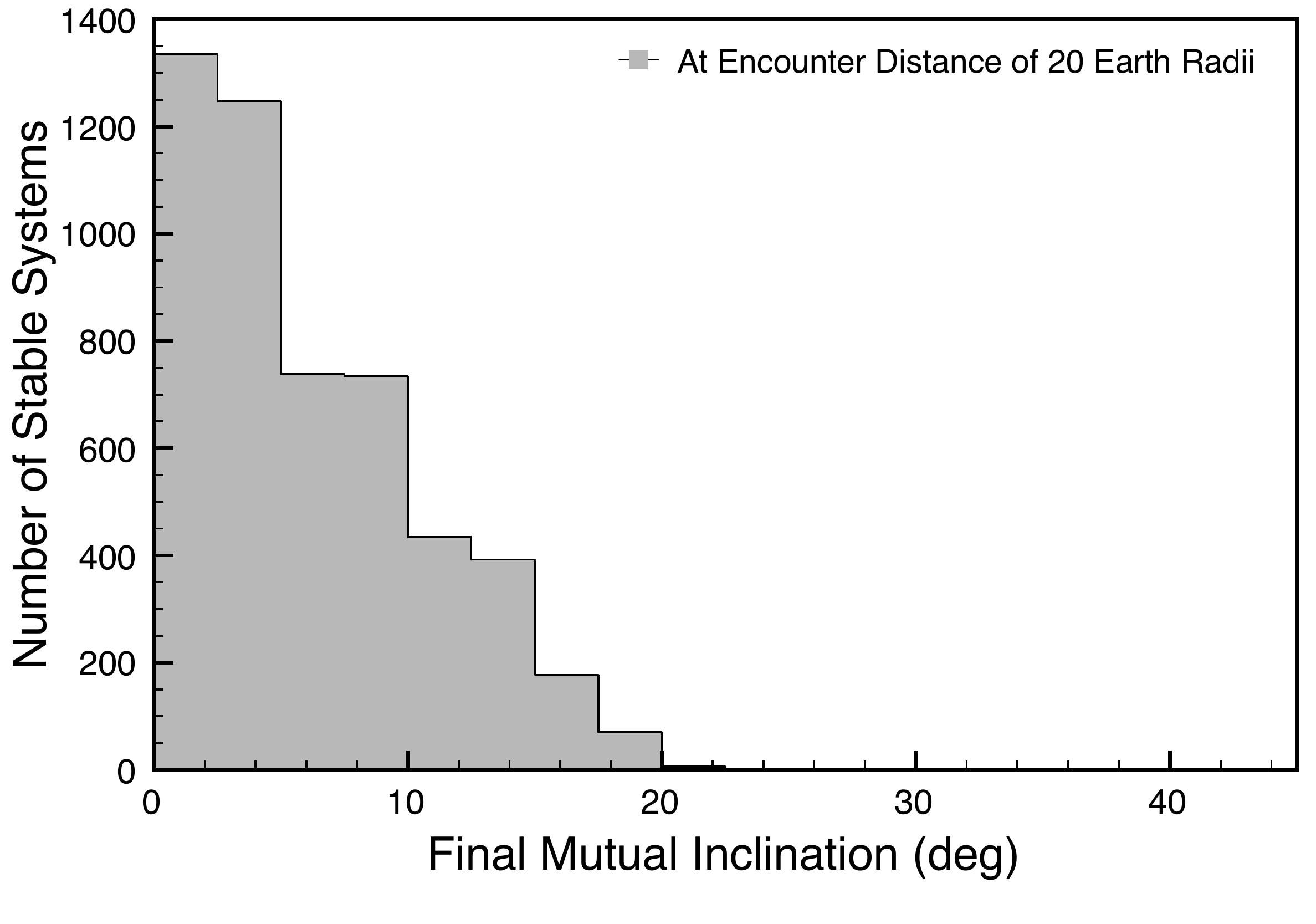}
\caption{2001 SN263: These histograms show the distribution of final eccentricities (top) and final mutual inclinations (bottom) for stable systems after close approaches of 20 $R_{\Earth}$. \label{snflyby2}} 
\end{figure}

\begin{figure}[bhtp]
	\centering
	\includegraphics[scale=0.35]{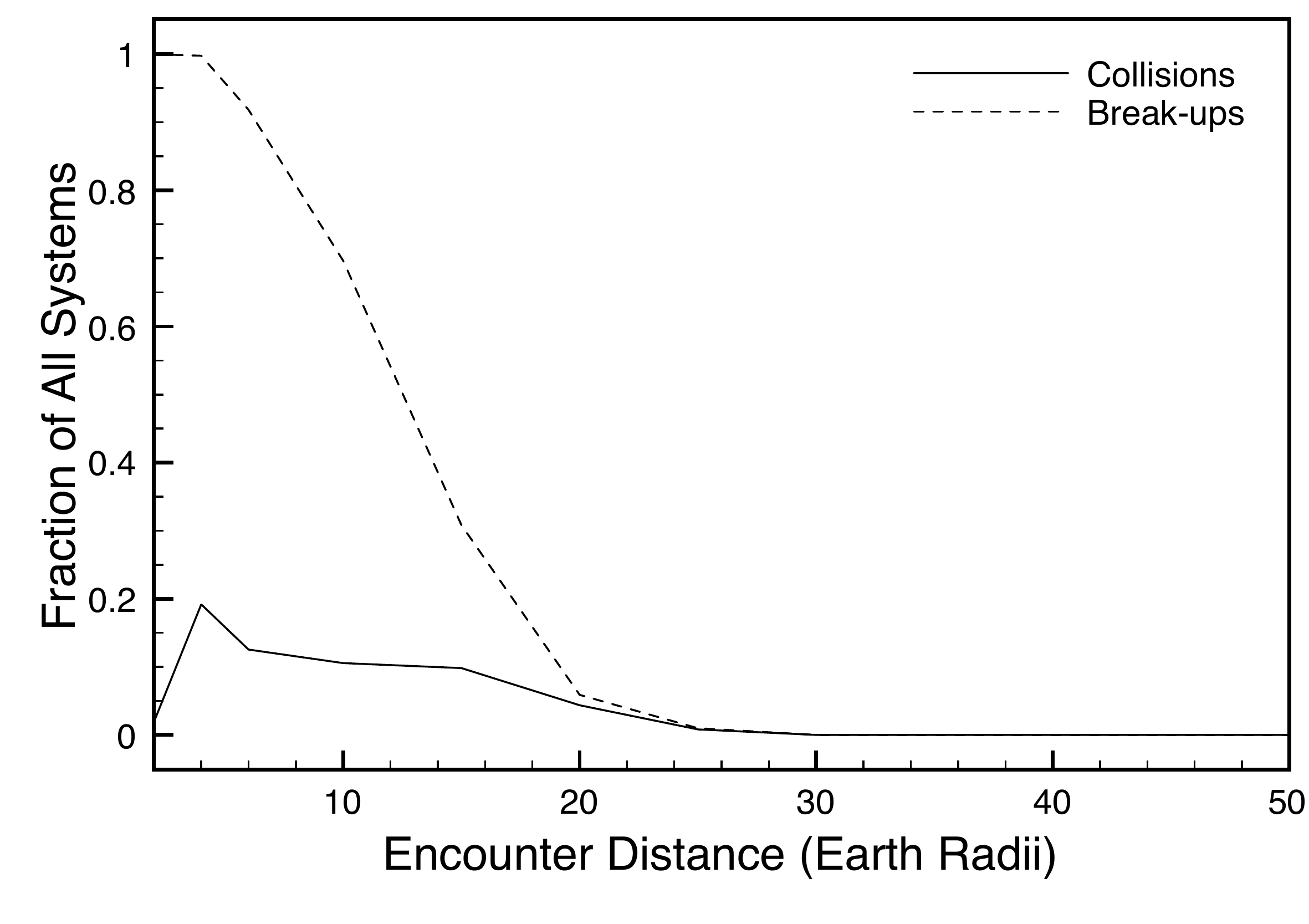}
	\includegraphics[scale=0.35]{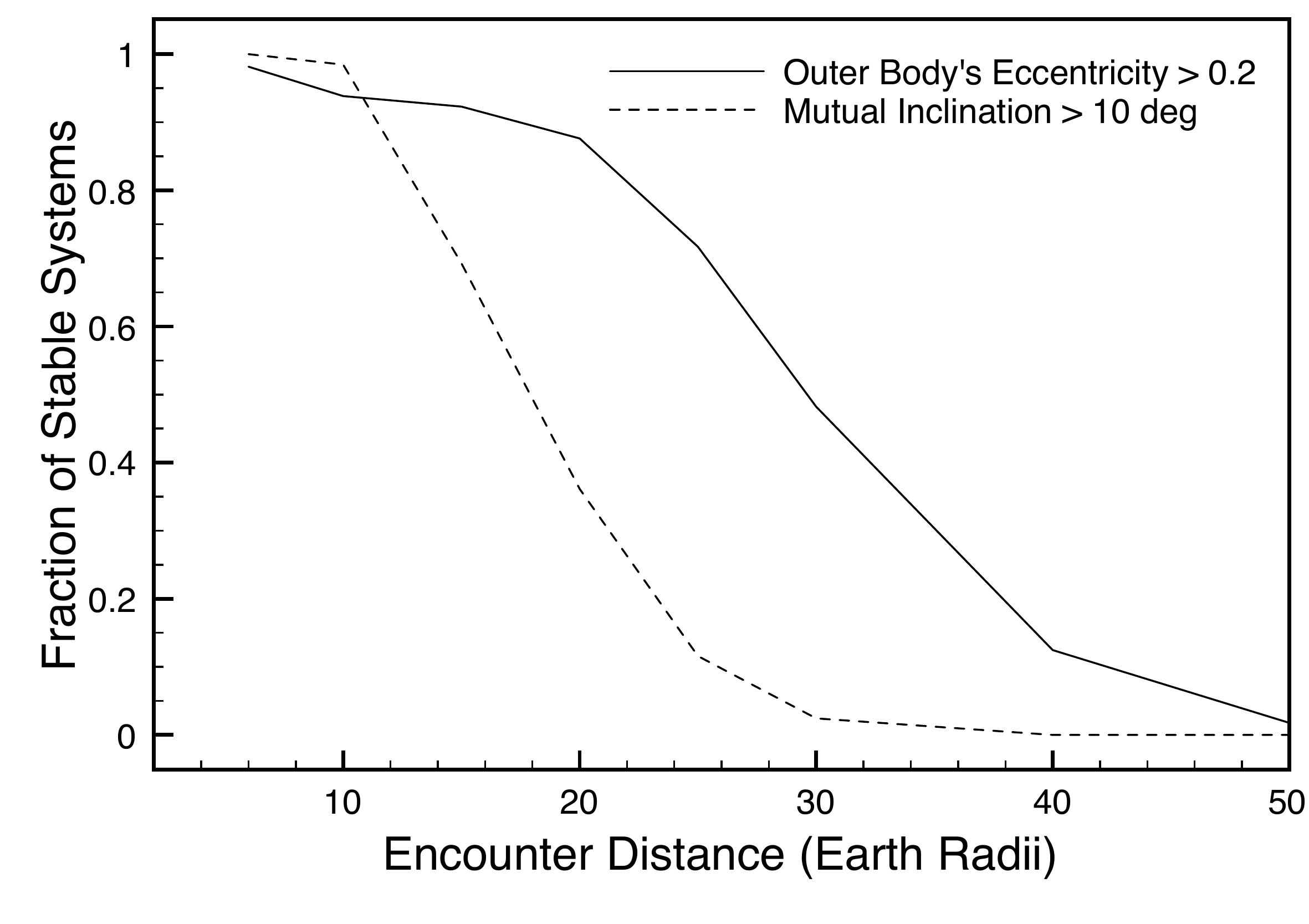}
\caption{1994 CC: These figures show the close encounter statistics for the onset of instability (top) and excitation (bottom). The top diagram shows the fraction out of all systems; the bottom diagram shows the fraction out of only stable systems, defined as those with no collisions and ejections. At encounter distances $\lesssim$ 5 $R_{\Earth}$, there are no stable systems left and as a result, we do not have excitation statistics for those very close encounters. \label{ccflyby1}} 
\end{figure}

\begin{figure}[tbtp]
	\centering
	\includegraphics[scale=0.35]{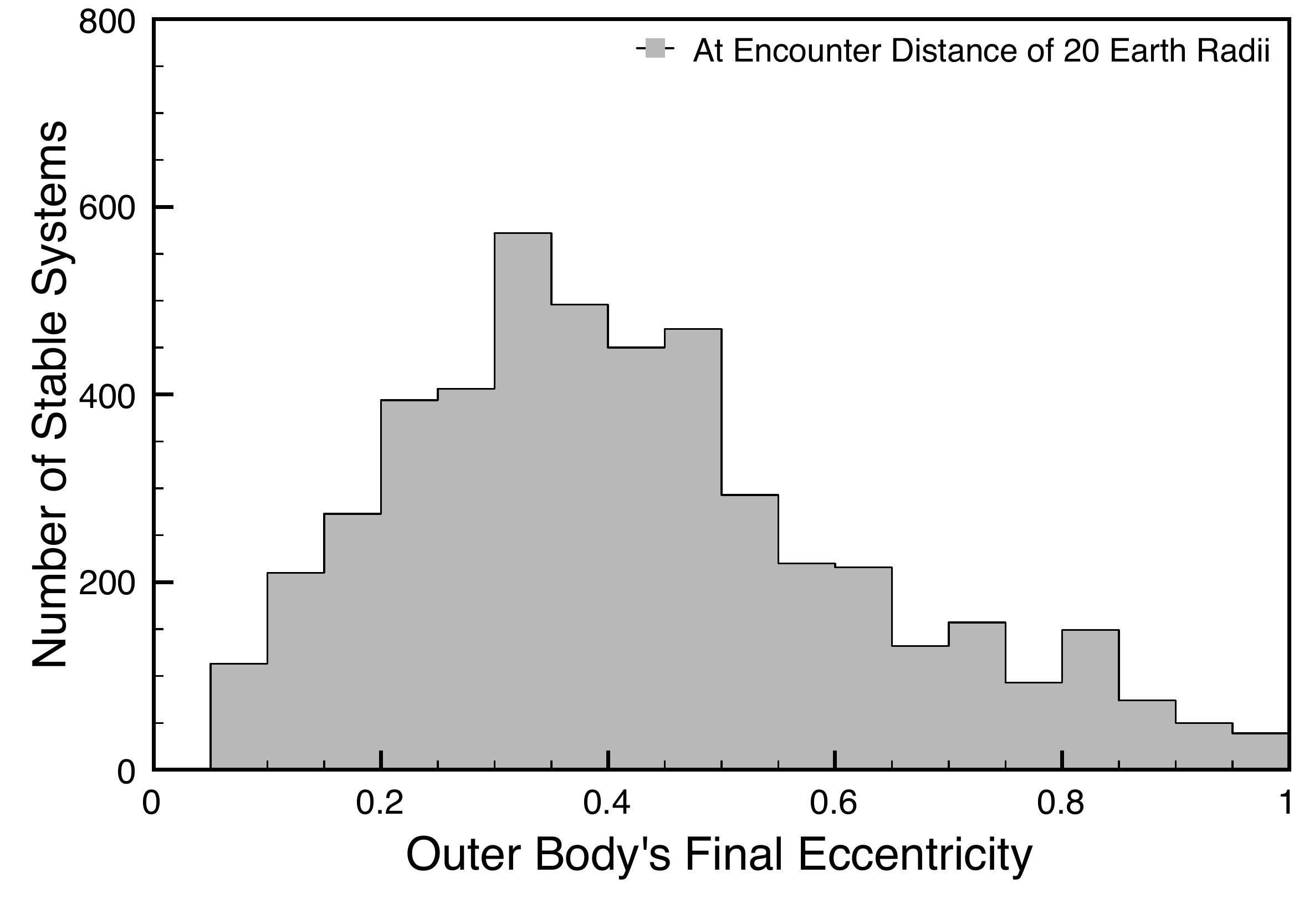}
	\includegraphics[scale=0.35]{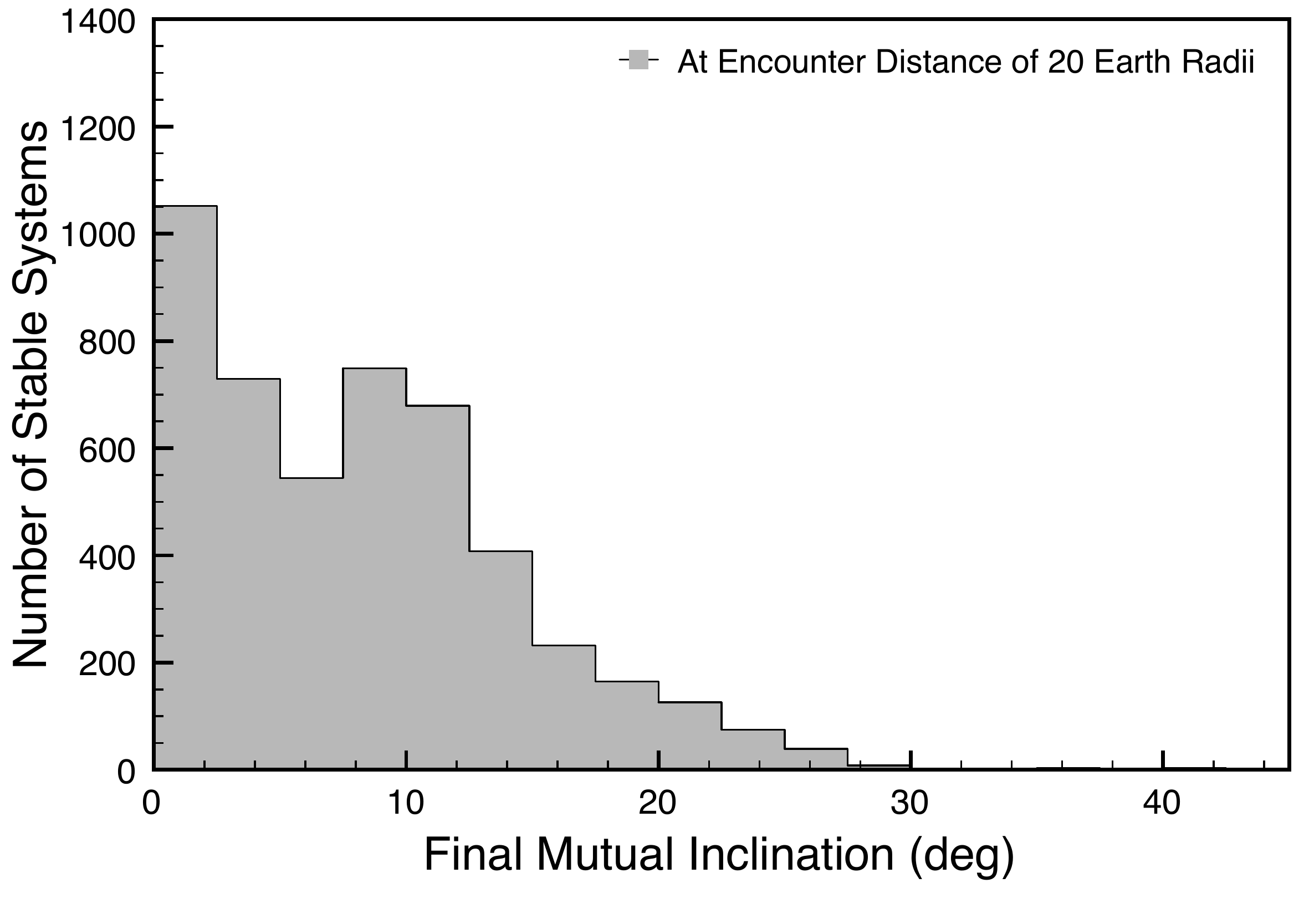}
\caption{1994 CC: These histograms show the distribution of final eccentricities (top) and final mutual inclinations (bottom) for stable systems after close approaches of 20 $R_{\Earth}$. \label{ccflyby2}} 
\end{figure}

Our close encounter results are shown in Figures \ref{snflyby1}-\ref{ccflyby2}, where we looked for (a) collisions between any of the bodies (Alpha, Beta, Gamma, and Earth-sized perturber) or (b) break-up of the system defined as ejection of either the inner or outer body, or both. For stable systems in which there were neither collisions nor break-ups, we examined cases where there were (c) excited eccentricities in the outer body of at least $\sim$0.2 seen in our orbit fitting for 1994 CC or (d) mutual inclinations greater than 10 degrees. The final eccentricities and inclinations were calculated from the mean of the osculating elements during the final few orbital periods. 

It is clear from Figures \ref{snflyby1}-\ref{ccflyby2} that 1994 CC is more easily disrupted and excited than 2001 SN263, which agrees with our earlier calculation of $a_{\rm satellite}/r_{\rm Hill}$ in Section 2. During close approaches with Earth, the outer body was easily excited to eccentricities of at least 0.2 as far away as encounter distances of $\sim$40 $R_{\Earth}$ for 2001 SN263 and $\sim$50 $R_{\Earth}$ for 1994 CC. The orbital planes of the satellites gained mutual inclinations of at least 10 degrees starting at encounter distances of $\sim$25 $R_{\Earth}$ for 2001 SN263 and $\sim$30 $R_{\Earth}$ 1994 CC. Unbound systems due to break-up/ejection scenarios started occurring at $\sim$20 $R_{\Earth}$ encounter distances for 2001 SN263 and $\sim$25 $R_{\Earth}$ for 1994 CC. The distribution of the outer body's final eccentricity and final mutual inclination between satellite orbits at a close encounter distance of 20 $R_{\Earth}$ are shown for 2001 SN263 (Figure \ref{snflyby2}) and 1994 CC (Figure \ref{ccflyby2}). We note that in Figures \ref{snflyby1} and \ref{ccflyby1}, at very close encounter distances (less than 5 $R_{\Earth}$) there are few stable systems left due to a high rate of ejections. As a result, for very close encounter distances the statistics shown for fraction of stable systems with excited eccentricities and mutual inclinations are more uncertain.

From these hyperbolic flyby simulations, we found that close planetary encounters could affect the orbits at distances as large as 50 $R_{\Earth}$. Such close approaches to Earth cannot at present happen, based on the current value of the minimum orbital intersection distance (MOID): $\sim$1190 $R_{\Earth}$ for 2001 SN263 and $\sim$380 $R_{\Earth}$ for 1994 CC (see Table \ref{moid}).  The MOID is the minimum separation between the osculating ellipses of the orbits of two bodies, without regard to position of the bodies in their orbits (Sitarski 1968); it remains valid as long as the osculating elements approximate the actual orbits. Over time, these elements will change, and it is likely that the observed high eccentricity and mutual inclinations were acquired during planetary encounters at a time when the MOID was lower and allowed for closer approaches.

\def\arraystretch{1.4}
\begin{deluxetable}{l r r}
\tablecolumns{3}
\tablecaption{Minimum Orbital Intersection Distance (MOID) \label{moid}}
\startdata
\hline \hline
& 2001 SN263 & 1994 CC \\
\hline
Mercury MOID (AU) & 0.695 & 0.508 \\
Venus MOID (AU) & 0.320 & 0.233 \\
Earth MOID (AU) & 0.0506 & 0.0162 \\
Mars MOID (AU) & 0.169 & 0.112
\enddata
\tablenotetext{}{This table shows the current minimum orbital intersection distance (MOID) between the triple-asteroid systems and the terrestrial planets using orbital elements that are valid at MJD 54509.0 (2001 SN263) and MJD 54994.0 (1994 CC).}
\end{deluxetable}

We can calculate a rough estimate for how often close planetary encounters occur for generic near-Earth systems (Chauvineau et al. 1995):
\begin{equation}
	t \sim \dfrac{2\tau_{\rm coll}R_{\Earth}^2}{b^2}
\end{equation}
where $t$ represents the time between encounters up to an impact parameter $b$ and $\tau_{\rm coll}$ is the asteroid's lifetime against impact with the terrestrial planets. For Earth, the average interval between impacts is $\sim$4 $\times$ $10^6$ and $\sim$3 $\times$ $10^5$ years for objects similar in size to 2001 SN263 and 1994 CC, respectively (Stuart \& Binzel 2004). There are approximately $\sim$150 near-Earth asteroids with a size comparable to 2001 SN263 and $\sim$2000 for 1994 CC, and accordingly, we estimate $\tau_{\rm coll} \sim$ 6 $\times$ $10^8$ years for both 2001 SN263 and 1994 CC. 

In our simulations with planetary encounters, break-up of the asteroid triples occurred in 50$\%$ of our systems at impact parameters $b \sim$ 10$R_{\Earth}$ for 2001 SN263 and $b \sim$ 13$R_{\Earth}$ for 1994 CC. Thus, we calculate $t$ to be $\sim$12 Myrs for 2001 SN263 and $\sim$7 Myrs for 1994 CC, which represent the estimated lifetimes of the systems due to such scattering encounters. These timescales are comparable to the $\sim$10 Myr dynamical lifetimes calculated by Gladman et al. (1997). A similar calculation for the time interval between planetary encounters that could excite 1994 CC Gamma's eccentricity using $b \sim$ 30$R_{\Earth}$ (when 50$\%$ of our simulations showed excited eccentricities of at least 0.2) yields $\sim$1 Myrs. As a result, close planetary encounters that occur on million-year timescales can reproduce the observed eccentricity and inclinations. These generic timescales could be improved with encounter calculations for specific near-Earth asteroids, such as those performed by Nesvorny et al. (2010).

We note that while close planetary encounters can explain the excited eccentricity and mutual inclinations in near-Earth triple systems like 2001 SN263 and 1994 CC, different processes are required for main belt asteroids and those trans-neptunian objects that have not experienced strong planetary scattering events.

\section{Conclusion}

In this work, we found dynamical solutions for two triple systems, 2001 SN263 and 1994 CC, where we have derived the orbits, masses, and Alpha's $J_2$ gravitational harmonic using full N-body integrations. We used range and Doppler data from Arecibo and Goldstone to solve this non-linear least-squares problem with a dynamically interacting three-body model that provided an excellent match to our radar observations. Given the three-body nature of these systems, we also measured the precession rates of the apses and nodes, and compared them to our corresponding analytical expressions from $J_2$ and secular contributions. No resonant arguments were found to be librating in either triple system.

For both systems, we detected significant mutual inclinations (2001 SN263: $\sim$14 deg, 1994 CC: $\sim$16 deg) between the orbital planes of Beta and Gamma. We also found a nonzero orbital eccentricity ($\sim$0.2) for 1994 CC's outer body, Gamma. The eccentricity and inclination damping timescales are long, suggesting that both systems are in excited states. We investigated excitation mechanisms that could explain the observed orbital configurations, including Kozai and evection resonances, mean-motion resonance crossings, and close encounters with the terrestrial planets. Close encounters that occur on million-year timescales can reproduce the observed mutual inclinations in both systems and 1994 CC Gamma's eccentricity.

\newpage
\section*{Acknowledgements}
We thank John Chambers for his assistance with the N-body integrator, \verb MERCURY , and Bill Bottke for useful discussions regarding close encounters. We are also grateful for the anonymous review that improved the manuscript. The Arecibo Observatory is part of the National Radio Astronomy and Ionosphere Center, which is operated by Cornell University under a cooperative agreement with the National Science Foundation. Some of this work was performed at the Jet Propulsion Laboratory, California Institute of Technology, under contract with the National Aeronautics and Space Administration (NASA). The paper is based in part on work funded by NASA under the Science Mission Directorate Research and Analysis Programs:  J.F. and J.L.M. were partially supported by the NASA Planetary Astronomy grant NNX09AQ68G.

\bibliographystyle{apj}

\end{document}